\documentclass[journal]{IEEEtran}
\ifCLASSINFOpdf
\else
\fi
\usepackage{float}
\usepackage{times}  
\usepackage{helvet} 
\usepackage{courier}  
\usepackage[hyphens]{url}  
\usepackage{amsmath,amsfonts}
\usepackage{algorithm}
\usepackage{algpseudocode}%
\usepackage{graphicx} 
\usepackage{graphicx}  
\newcommand{\citet}[1]{\citeauthor{#1} \shortcite{#1}}
\newcommand{\citep}{\cite}

\usepackage{placeins}
\usepackage{subcaption}
\usepackage{cite}
\usepackage{tcolorbox}
\usepackage{tabularx}
\usepackage{xcolor}
\tcbset{tab2/.style={colback=red!5!white,colframe=red!50!black,colbacktitle=red!40!white,
coltitle=black,center title}}

\usepackage{colortbl}

\begin{document}
%
\title{Multi-task Learning Approach for Modulation and Wireless Signal Classification for 5G and Beyond: Edge Deployment via Model Compression}


\author{\IEEEauthorblockN{Anu Jagannath and Jithin Jagannath\\}
\IEEEauthorblockA{Marconi-Rosenblatt AI/ML Innovation Lab\\
ANDRO Computational Solutions, LLC
New York, USA\\
Email: \{jjagannath, ajagannath\}@androcs.com}\\
\thanks{ACKNOWLEDGMENT OF SUPPORT AND DISCLAIMER:(a) Contractor acknowledges Government’s support in the publication of this paper. This material is based upon work supported by the US Army Contract No. W9113M-20-C-0067. (b) Any opinions, findings and conclusions or recommendation expressed in this material are those of the author(s) and do not necessarily reflect the views of the US Army.}
}


%


\maketitle

\begin{abstract}
Future communication networks must address the scarce spectrum to accommodate extensive growth of heterogeneous wireless devices. Efforts are underway to address spectrum coexistence, enhance spectrum awareness, and bolster authentication schemes. Wireless signal recognition is becoming increasingly more significant for spectrum monitoring, spectrum management, secure communications, among others. Consequently, \emph{comprehensive spectrum awareness on the edge} has the potential to serve as a key enabler for the emerging beyond 5G (fifth generation) networks. State-of-the-art studies in this domain have (i) only focused on a single task - modulation or signal (protocol) classification - which in many cases is insufficient information for a system to act on, (ii) consider either radar or communication waveforms (homogeneous waveform category), and (iii) does not address edge deployment during neural network design phase. In this work, for the first time in the wireless communication domain, we exploit the potential of deep neural networks based multi-task learning (MTL) framework to simultaneously learn modulation and signal classification tasks while considering heterogeneous wireless signals such as radar and communication waveforms in the electromagnetic spectrum. The proposed MTL architecture benefits from the mutual relation between the two tasks in improving the classification accuracy as well as the learning efficiency with a lightweight neural network model. We additionally include experimental evaluations of the model with over-the-air collected samples and demonstrate first-hand insight on model compression along with deep learning pipeline for deployment on resource-constrained edge devices. We demonstrate significant computational, memory, and accuracy improvement of the proposed model over two reference architectures. In addition to modeling a lightweight MTL model suitable for resource-constrained embedded radio platforms, we provide a comprehensive heterogeneous wireless signals dataset for public use.

\end{abstract}


%
\IEEEpeerreviewmaketitle

\section{Introduction}

\IEEEPARstart{S}{pectrum} sensing for comprehensive situational awareness will play an essential role in Beyond 5G (5th Generation) networks. Some of the direct applications include promoting coexistence in the unlicensed spectrum bands, automatic signal recognition for advanced physical layer security, intrusion detection, among others. The need for an advanced wireless signal recognition system in the present and forthcoming age of wireless communication where heterogeneous and dense wireless devices of civilian, commercial, military, and/or government domains share and contest for the scarce spectrum is critical. The Internet of Things (IoT) adoption and deployment scale is forecasted to grow at an unprecedented rate. To this end, the requirements of future communication networks are already set to support at least $10\times$ the device density (device/km$^2$) of 5G \cite{Ajagannath6G2020} to sustain diverse application domains - smart grid, smart city, holographic teleconferencing, industrial automation, etc. Such dense deployment is often accompanied by a plethora of security vulnerabilities owing to the various exposed threat surfaces. An automated signal recognition scheme for coordinated spectrum sharing, physical layer authentication, intruder detection, etc., is indispensable for secure communications in the Beyond 5G (B5G) networks.


Signal recognition involves extracting waveform descriptors such as the wireless standard (protocol), modulation format, and/or hardware intrinsic signatures, among others. Cooperative or coordinated spectrum sharing involves a preliminary signal sensing and identification process to distinguish the authorized, unauthorized, and rogue emitters in the vicinity contesting for the scarce spectrum. For instance, the coexistence of 5G New Radio Unlicensed (NR-U) devices with incumbents in the unlicensed spectrum is an actively studied topic by the industry as well as a study item in the 3GPP Release-17 working group \cite{rel17a,rel17b}. Another instance is the spectrum sharing in the 150 MHz of Citizens Broadband Radio Service (CBRS) in the 3.5 GHz radio band \cite{cbrs}. 

Signal recognition is composed of subtasks such as modulation recognition \cite{Zhou_AMC_survey,Jagannath19MLBook}, wireless standard (protocol) determination \cite{SignalRecogSurvey}, RF fingerprinting \cite{Ajagannath2022ComST2022}, etc. However, tackling these subtasks as a joint problem could benefit from the similarities shared across these tasks. State-of-the-art in this realm has focused on studying these subtasks independently and only considers either common communication waveforms \cite{AlexGoogleNet_AMC,Jagannath18ICC,oshea2018,ICAMCNet} or radar signals \cite{radarsignalsbinary}. Recently, however, since spectrum sharing between incumbents and/or authorized/unlicensed devices is gaining momentum, considering heterogeneous waveform families - radar and/or communication waveforms - would be essential. To this end, we extend our previous work \cite{AJagannath21ICC} to design and comprehensively evaluate the wireless signal recognition problem with subtasks - modulation and wireless standard recognition - jointly in a multitask setting for radar as well as communication waveforms using synthetic and over-the-air (OTA) datasets. Further, most of the existing works \cite{AlexGoogleNet_AMC,ICAMCNet,oshea2018,wirelessInterference,wirelesstech} focus on designing a \emph{deep} neural network to accomplish the task at hand, be it modulation or protocol classification, without considering the deployment platform capabilities. We claim that such a design methodology which overlooks the target platform capabilities or are built under the assumption of a powerful computational unit for model deployment limits the IoT spectrum awareness capabilities as required by future communication standards. Therefore, in this work, we also put forth a suitable deep learning pipeline with special emphasis on resource-constrained edge deployment.

To the best of our knowledge, our proposed MTL model is the first in the deep learning for wireless communication domain that introduces MTL to solve challenging multiple waveform characterization tasks simultaneously. Further, MTL architecture inherently generalizes better with more number of tasks since the model learns shared representation that captures all tasks. Hence, additional signal classification or regression tasks can be included in the model without significantly diminishing its performance. In order to elucidate the overall contributions of this extended work, we enlist the key differences of this version from our preliminary work \cite{AJagannath21ICC}.

\begin{itemize}
    \item In this extended version, we have revised and elaborated the Related Works (section II) to better illustrate the evolution of modulation classification and the need for MTL from an edge deployment perspective. In this work, we demonstrate the significant computational gain and lesser memory requirement with the MTL architecture in contrast to other reference architectures.
    \item We walk through and elaborate the MTL design ethos which considers edge deployment from its design inception. We present a comprehensive insight into the design methodology with insight on the computational requirements, model convergence, and accuracy in section V-C.
    \item To explicitly enable the reader to better understand the benefits of MTL, we have clearly enlisted it in Section III as well as introduces the training procedure for MTL. 
    \item In this extended version, we have included waveform visualizations from the OTA collection for better comprehension. Additionally, the challenging OTA collected dataset (RadComOta) is made available to benefit the wider research community \cite{data}. The dataset collection presents an in-the-wild OTA scenario with other unavoidable emitters in the vicinity. Such a unique combination of radar and communication waveform dataset annotated to suit multiple tasks are not available to date. We would like to emphasize here that the dataset can be utilized for two single tasks as well - modulation and wireless standard recognition separately.
    \item Further, we demonstrate the performance of the deduced lightweight MTL model on an experimental OTA testbed and exhibit a 90\% signal classification and 82.5\% modulation classification accuracy at very low signal powers. The performance is contrasted with two benchmark models and shown to outperform significantly in terms of computation, accuracy, and memory. We also elaborate the OTA data collection environment and demonstrate the received signal strengths to elucidate the experimental settings to the reader specifically in section VI.
    \item Finally, we present a deep learning pipeline which considers model compression as a critical step for deploying \emph{neural network-assisted spectrum awareness at the edge} in the beyond 5G communication networks. We have included instructional firsthand information on performing model compression. Accordingly, we demonstrate the model size reduction ($11.8\times$) by quantizing the MTL model with negligible accuracy loss. This discussion is included in section VII.
\end{itemize}

\section{Related Works}

\begin{figure}[h]
\centering
\includegraphics[width=0.99\columnwidth]{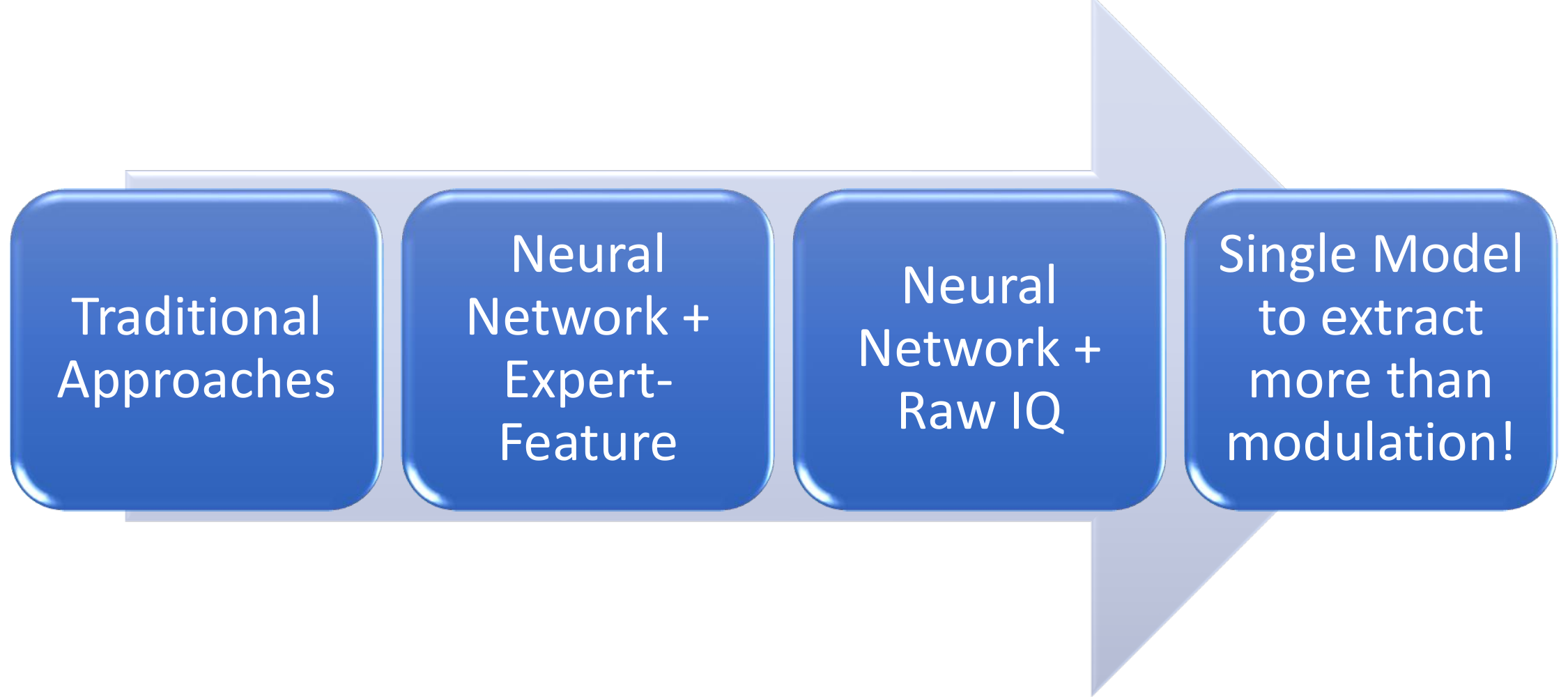} 
\caption{Evolution of AMC Approaches}
\label{fig:Evolution}
\end{figure}

Machine learning is becoming a key enabler for several aspects of the next-generation (B5G) of wireless communication systems and RF signal analysis  \cite{Zhou_AMC_survey,JagannathAdHoc2019,SignalRecogSurvey, Ajagannath6G2020}. One of the most common tasks of wireless signal recognition is automatic modulation classification (AMC) whereby the modulation type of the RF waveform is predicted by the receiver. AMC has now been studied over several years and has therefore evolved as the time has progressed. We first begin by looking at this evolution of AMC as depicted in Fig. \ref{fig:Evolution}.

\textbf{Traditional Approaches:} AMC was broadly divided into two categories; (i) feature based \cite{Feature_1, Cumulant_Chang, FB_Sudhan_2017, Han_1} and (ii) likelihood based methods \cite{ML_Hameed, LB_Jianping, Dis_AMC, ML_Su_2,Varshney_1,Varshney_2}. While likelihood based approaches provide optimal performance in the Bayesian sense, they are often computationally demanding and not suited for deployment at the edge where most application for spectrum awareness lies \cite{Jagannath_CCWC, Foulke}. On the other hand, feature based classifiers are computationally efficient and can provide near optimal performance if designed carefully. The requirement of being ``carefully designed" is the key caveat here. There has also been attempts to combine the benefits of both in hybrid approaches \cite{Jagannath17CCWC}. It is often possible to design the classifier to work extremely well under certain assumption in simulations or laboratory setting but fail when the operational environment changes. In other words, it is important for the classifiers to generalize well to various operating scenarios.

Since the problem structure of feature based classifiers fits extremely well with the recently revitalized supervised machine learning, it was inevitable for these techniques to be leveraged for AMC. Therefore, in recent years, different machine learning techniques have been employed to determine the modulation format of the unknown signal via classification. During the initial stages of applying supervised learning for AMC, feature-engineered methodology was adopted as opposed to raw inphase-quadrature (IQ) samples. 
This includes the use of support vector machines (SVMs) \cite{SVM} and ANNs \cite{ANN2,ANN,ANN3}.  In \cite{ANN2}, the authors use a multilayer perceptron (MLP) to classify twelve different modulation formats with high accuracy over a wide range of signal-to-noise ratio (SNR) values. In \cite{ANN}, the authors use six features and evaluate two different ANN architectures trained by the backpropagation method using the standard gradient descent (GD) learning algorithm. 
Similarly, in \cite{ANN3}, eight modulation schemes have been shown to be successfully classified with high accuracy in low SNR conditions. All these studies are limited to simulations and not evaluated on actual hardware. In \cite{Foulke}, authors described challenges while transitioning from simulation to hardware implementation. Overall, due to the assumptions and unanticipated signal distortions that are overlooked during simulations, OTA performance of AMC techniques may experience degradation in real deployments. Therefore, valuing practical relevance, we demonstrate the performance of the MTL model with OTA captured waveforms. 

\textbf{Neural network with expert-feature:} The superior feature extraction capability of convolutional neural networks (CNNs) as opposed to ANNs led to several works utilizing CNNs for modulation or signal classification \cite{AlexGoogleNet_AMC, vhf_amc,ICAMCNet,oshea2018,wirelessInterference,wirelesstech}. The authors of \cite{AlexGoogleNet_AMC} evaluated the performance of convolutional neural networks (CNNs) in predicting modulation formats on a dataset comprising eight classes. Here, the authors adopted GoogLeNet \cite{googlenet} and AlexNet \cite{alexnet} architectures which were fed constellation images as input. However, the models exhibited sensitivity to image preprocessing factors such as image resolution, cropping size, selected area, etc., and achieved an accuracy below 80\% at 0 dB signal-to-noise ratio (SNR). We claim here that this could be due to employing architectures designed to tackle computer vision problems rather than for the RF application. A classification accuracy of 98\% on seven modulations was demonstrated on a Universal Serial Radio Peripheral (USRP) software-defined radio (SDR) testbed with a feed-forward feature-based neural network \cite{Jagannath18ICC}. A seven class radar waveform recognition with a CNN architecture that ingests time-frequency images was studied in \cite{radar_recog}. A seven class modulation recognition accuracy of 95\% at SNRs above 2 dB was attained with a CNN utilizing cyclic spectrum images \cite{vhf_amc}. All of these works rely on handcrafted features to train the neural network which limits the generalization capability of the network as it could have from raw IQ samples. Therefore, in this work we resort to extracting spatial features from raw IQ samples instead of hand-engineered feature sets.

\textbf{Neural network with raw (IQ or real-valued) samples:} 

In \cite{vtcnn}, the authors use 128 raw IQ samples (per example) to classify 11 communication modulation schemes with  a CNN architecture referred to as VTCNN. The network was shown to achieve a classification accuracy in the range of 70\% - 75\% for signals with 18 dB SNR. Similarly, in \cite{drcnn} 128 raw IQ samples were utilized to classify 8 modulations with another CNN architecture (DRCNN) which achieved above 95\% accuracy for signals exceeding 2 dB. The authors use 2000 point real-valued samples in \cite{XuLiWang2019} to classify $5$ communication waveforms with a CNN architecture. Although the model achieves a $100\%$ accuracy it considers very limited number of waveforms of the same carrier frequency and bandwidth. In contrast, our work considers and captures a diverse and comprehensive setting with respect to sampling rate, carrier frequency, multipath propagation effects, among others which is more representative of a realistic scenario. The authors of \cite{ICAMCNet} trained a CNN with 1024 raw IQ samples per input to classify 11 modulations and achieved an accuracy of 83.4\% at 18 dB. In \cite{oshea2018}, a modified ResNet architecture was trained to perform a 24 modulation class predictions with 95.6\% accuracy at 10 dB by learning from raw IQ samples. Note here that these works are adopting deeper architectures for singular task unlike the proposed MTL model. A single modulation classification task for communication waveforms was further split into subtasks in a multi-task learning (MTL) setting in \cite{mtlmod}. However, it is performing only a single waveform characterization task - modulation recognition. These works studied single task modulation classification tasks on only communication waveforms. On the contrary, our proposed MTL model can attain two waveform characterization tasks - modulation and signal (protocol/standard) recognition - on the radar as well as communication waveforms with a single model. Further, in this work, we demonstrate the MTL model performance with OTA captured waveforms and achieve a 82.5\% modulation accuracy and 90\% signal classification accuracy at the lowest transmission gain (0 dB).

Another subtask of wireless signal recognition is signal (wireless protocol) recognition which involves identifying the wireless standard with which RF waveform is generated. The authors of  \cite{wirelessInterference} studied wireless interference detection involving three wireless standards - IEEE 802.11 b/g, IEEE 802.15.4, and IEEE 802.15.1 - occupying different frequency channels grouped into 15 different classes. In a similar sense, \cite{wirelesstech} adopted a CNN architecture to address the spectrum crunch in the industrial, scientific, and medical (ISM) band by identifying seven classes belonging to Zigbee, WiFi, Bluetooth, and their cross-interferences. However, the model required operation in a high SNR regime for a 93\% accuracy on a singular task.

\textbf{Dataset:} Deep learning has made significant strides in the field of computer vision \cite{DL_CV,alexnet}, natural language processing \cite{nlp}, speech recognition \cite{DL_speech}, etc. However, its application in the field of wireless communication is still in its early stages. The recent application of deep learning in wireless communication is starting to witness rapid advancements in the field of wireless resource allocation, modulation recognition, intelligent receiver designs, etc., \cite{JagannathAdHoc2019}. The comparatively slower pace of applied deep learning in wireless communication in contrast to other domains can be in part attributed to the lack of available large scale datasets for the diverse wireless communication problems.  
It is well known that datasets are the fuel on which machine learning thrives. In our literature search \cite{oshea2018,crawdad,OSheaCNN2,genesys_powder}, we have not yet come across any dataset that has been curated for multi-task learning architectures or that contains both communication and radar waveforms. Therefore, in this effort, we had to generate elaborate datasets - RadComAWGN, RadComDynamic, and RadComOta - comprising communication and radar signals with appropriate annotations that can be used for multi-task and single task approaches (modulation and signal classification separately). To benefit future research in this field, the RadComDynamic and RadComOta have been made public \cite{data}. Consequently, this dataset can be availed by the scientific community to study wireless modulation and signal classification separately or jointly.

\section{Wireless Multi-task Learning}
Multi-task learning (MTL) is a neural network paradigm for inductive knowledge transfer which improves generalization by learning shared representation between related tasks. MTL improves learning efficiency and prediction accuracy on each task in contrast to training an STL model for each task \cite{mtl_95}. MTL has been applied to natural language processing (NLP) and computer vision extensively. In \cite{mtl_2019_1}, an MTL framework to jointly perform medical named entity recognition and normalization from medical literature was proposed. The parallel multi-task architecture is setup to perform explicit feedback between tasks allowing conversion of hierarchical tasks. Another NLP work \cite{mtl_2019_2} proposed an MTL classifier to perform sarcasm and sentiment classification. An MTL approach to perform parallel regression and classification tasks on a monocular input image whereby the tasking weights are learned is proposed in \cite{mtl_2018_CVPR}. Another study in \cite{mtl_2015_CV} considered scene images from different resolutions as related tasks that could be jointly learned with an MTL model.

Although MTL has been widely applied in NLP and computer vision, its popularity in the wireless communication domain is yet to gain traction. To the best of our knowledge, this is the first work that applies MTL to jointly tackle waveform characterization tasks (i.e. automatic modulation classification and signal classification). We exploit the mutual relation between the tasks in applying them to an MTL setting. Specifically, we adopt a hard-parameter shared architecture \cite{Caruana1993} where there exists a shared branch (shares hidden layers with all tasks) and task-specific branches. It was shown in \cite{Baxter1997} the hard-parameter shared model reduces overfitting risk by the order of the number of tasks. The model extracts a shared representation that captures all of the tasks consequently improving the generalization capability. Therefore, the inclusion of more tasks will improve the model's learning efficiency. To this end, in this work, we consider two related waveform characterization tasks - modulation and signal classification - that can benefit from each other with the hard-parameter shared model. We would like to articulate here the benefits of MTL in contrast to training STL models per task:
\begin{enumerate}
    \item MTL solves a set of tasks jointly rather than independently which in theory comes with benefits such as reduced training and inference times, increased data efficiency, and improved prediction accuracy \cite{pmlr-v119-standley20a}.
    \item Reduced computational and storage resources due to a single MTL architecture as opposed to requiring multiple architectures each of which is optimized for their own individual tasks.
    \item Extensible architecture enables inclusion of related tasks without significant restructuring. For instance, few related tasks that can be included in the future are frequency estimation, emitter identification (RF fingerprint extraction), etc. In other words, it is able to provide more actionable information regarding the spectrum of interest with a single model.  
    \item Synthesizes a generalized architecture that does not overfit to a particular task. This can be attributed to the model's ability in learning shared representation. In other words, if a particular task A has a feature set $\mathcal{\alpha}$ that is related to it in a complex manner, Learning $\mathcal{\alpha}$ through A will be challenging. On the other hand, this feature set $\mathcal{\alpha}$ could be related to another task B in a less complex manner allowing the MTL model to learn the feature set through B. It can be said that the shared branch of the MTL allows the model to eavesdrop on the shared features of both tasks.
    \item It is known that wireless domain unlike computer vision has only scarce data available for training neural networks. MTL models are good at handling scarce data. For example, consider two tasks A and B whereby one of the tasks say A has lesser data than the other, training a STL for learning task A would be difficult. However, MTL can handle such uneven data distribution fairly well by allowing task A to benefit from task B.
\end{enumerate}

The proposed MTL model will classify the input waveform as belonging to a particular modulation and signal class concurrently. Both classification tasks are optimized with categorical cross-entropy losses denoted by $\mathcal{\ell}_{m}$ and $\mathcal{\ell}_{s}$ respectively. The joint MTL loss ($\mathcal{\ell}_{mtl}$) function is represented as a weighted sum of losses over the two tasks as in equation (\ref{eq:mtlloss}).
\begin{equation}
\ell_{mtl}(\Theta_{sh},\Theta_{m},\Theta_{s})=w_{m}\ell_{m}(\Theta_{sh},\Theta_{m}) + w_{s}\ell_{s}(\Theta_{sh},\Theta_{s})
    \label{eq:mtlloss}
\end{equation}
Here, $w_{m}$ and $w_{s}$ denote the weights over the task specific losses and \{$\Theta_{sh}, \Theta_{m}, \Theta_{s}$\} denote the learnable parameters of the shared and task-specific branches. The joint multi-task loss is parameterized by the shared as well as task-specific parameters. The MTL training is denoted as the optimization in equation (\ref{eq:mtlopt}).
\begin{align}
    \Theta_{sh}^{*},\Theta_{m}^{*},\Theta_{s}^{*} = \underset{\Theta_{sh},\Theta_{m},\Theta_{s}}{\arg\min}\ell_{mtl}(\Theta_{sh},\Theta_{m},\Theta_{s}) \label{eq:mtlopt}
\end{align}
The MTL optimization aims to tune the network parameters such as to minimize the overall task loss as shown in Algorithm 1. In this way, the model jointly learns to perform simultaneous predictions at a time.

\begin{algorithm}
\caption{Backpropagation to train MTL model}
\begin{algorithmic}
\State Initialize network weights $\Theta_{sh},\Theta_{m},\Theta_{s}$.
\State Initialize task weights $w_{m}$ and $w_{s}$.
\For{\texttt{epoch = 1 to MAX\_EPOCHS}} 
\For{\texttt{steps = 1 to STEPS}}
\State \textbf{Input} batch $\mathbf{x}$ and \textbf{Compute} MTL joint loss \State $\ell_{mtl}(\Theta_{sh},\Theta_{m},\Theta_{s})$ [standard forward pass]
\State \textbf{Compute} gradients $\nabla \ell_{mtl}(\Theta_{sh},\Theta_{m},\Theta_{s})$
\State \textbf{Update} weights \State $\Theta_{sh}^{*},\Theta_{m}^{*},\Theta_{s}^{*}\longleftarrow\Theta_{sh},\Theta_{m},\Theta_{s}$ [standard backward \State pass]
\EndFor
\State Early stopping monitor to track model convergence - \State Stop training once model stops learning (starts to di\State verge)
\EndFor
\end{algorithmic}
\end{algorithm}

\begin{figure}[h]
\centering
\includegraphics[width=0.8\columnwidth]{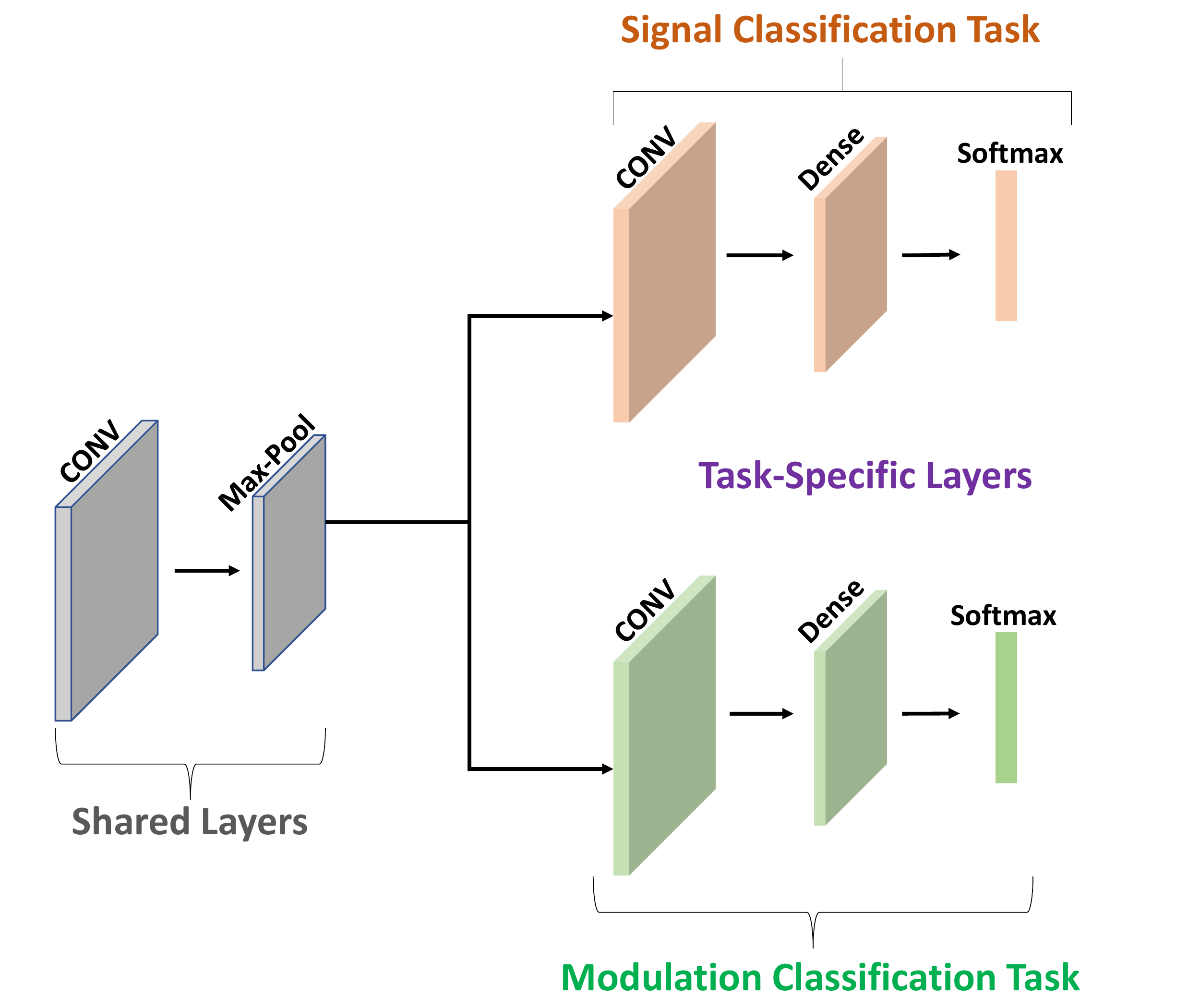} 
\caption{MTL architecture for spectrum awareness}
\label{fig:mtl}
\end{figure}

\textbf{MTL Network Architecture:} The hard parameter shared MTL architecture for wireless signal recognition is shown in Fig. \ref{fig:mtl}. The shared hidden layers are composed of convolutional and max-pooling layers. Each task-specific branch is comprised of convolutional, fully connected, and output softmax classification layers. The convolutional and fully-connected layers in the network adopt ReLU activation function.

The hyper-parameters such as convolutional kernel sizes, number of neurons in each layer, number of layers, task loss weights, etc., and their effects on the training performance and classification accuracies were studied in depth as elaborated in the upcoming sections. We train the network with Adam \cite{adam} gradient descent solver for 30 epochs with a patience of 5. The learning rate is set to 0.001. The architecture adopts batch normalization prior to ReLU activation. The dropout rate of shared layer is set to 0.25 and that of the task-specific branches are set to 0.25 and 0.5 in the convolutional and fully-connected layers respectively. Unless otherwise stated all the kernel sizes in the convolutional layers are $3\times3$ and max-pooling size is $2\times2$. We pass the model output through a softmax layer and sample from the resulting probability vector,
\begin{equation}
    p(\mathbf{y|f^{\Theta_i}(x)}) = \text{Softmax}(\mathbf{f^{\Theta_i}(x)})
\end{equation}
Here $\mathbf{i}$ indicates the corresponding task which could be modulation or signal classification, $\mathbf{x}$ and $\mathbf{f^{\Theta_i}(x)}$ denote the model input and output.
We implement our models in Keras with Tensorflow backend on an Ubuntu 18.04 VM running on an Intel Core i5-3230M CPU.

\begin{table}[b!]
    \centering
    \captionof{table}{Modulation and corresponding signal (protocol) classes}
    \label{tab:classes}
    \begin{tcolorbox}[tab2,tabularx={|m{4 cm}||m{5 cm}}]
      \textbf{Modulation Classes}   & \textbf{Signal Classes}  \\ \hline\hline
      BPSK   & SATCOM   \\ \hline
    ASK & Short-Range   \\ \hline
    AM-DSB \newline AM-SSB  & AM Radio \\\hline 
    GFSK   & Bluetooth   \\ \hline
    DSSS-CCK   & IEEE802.11bg   \\\hline
    DSSS-OQPSK   & IEEE802.15.4   \\ \hline
    FMCW (Radar)   & Radar-Altimeter (Radar)   \\  \hline
    PCW (Radar)   & Airborne-detection (Radar) \newline Airborne-range (Radar) \newline Ground mapping (Radar) \newline Air-Ground-MTI (Radar) \\ 
    \end{tcolorbox}{}
\end{table}

\begin{figure*}
\begin{minipage}[h]{0.4 \linewidth}
 \captionof{table}{RadComDynamic: Dynamic settings}\label{tab:dyn}
\small
\centering
    \begin{tcolorbox}[tab2,tabularx={|p{4.5 cm}||p{1.5 cm}}]
     \textbf{Dynamic Parameters}                                   & \textbf{Value}       \\ \hline\hline
Carrier freq. offset std. dev/sample        & 0.05 Hz            \\ \hline
Maximum carrier frequency offset                              & 250 Hz             \\ \hline
Sample rate offset std. dev/sample              & 0.05 Hz            \\ \hline
Maximum sample rate offset                                    & 60 Hz              \\ \hline
Num. of sinusoids in freq. selective fading        & 5                    \\ \hline
Maximum doppler frequency                                     & 2 Hz                 \\ \hline
Rician K-factor                                               & 3                    \\ \hline
Fractional sample delays comprising power delay profile (PDP) &$[0.2, 0.3, 0.1]$ \\ \hline
Number of multipath taps                                      & 5                    \\ \hline
Magnitudes corresponding to each delay in PDP         & $[1, 0.5, 0.5 ]$   \\ \hline
    \end{tcolorbox}{}
\end{minipage}
\hspace{0.5 cm}
\begin{minipage}[h]{0.55 \linewidth}
\centering
\includegraphics[width=.99 \columnwidth]{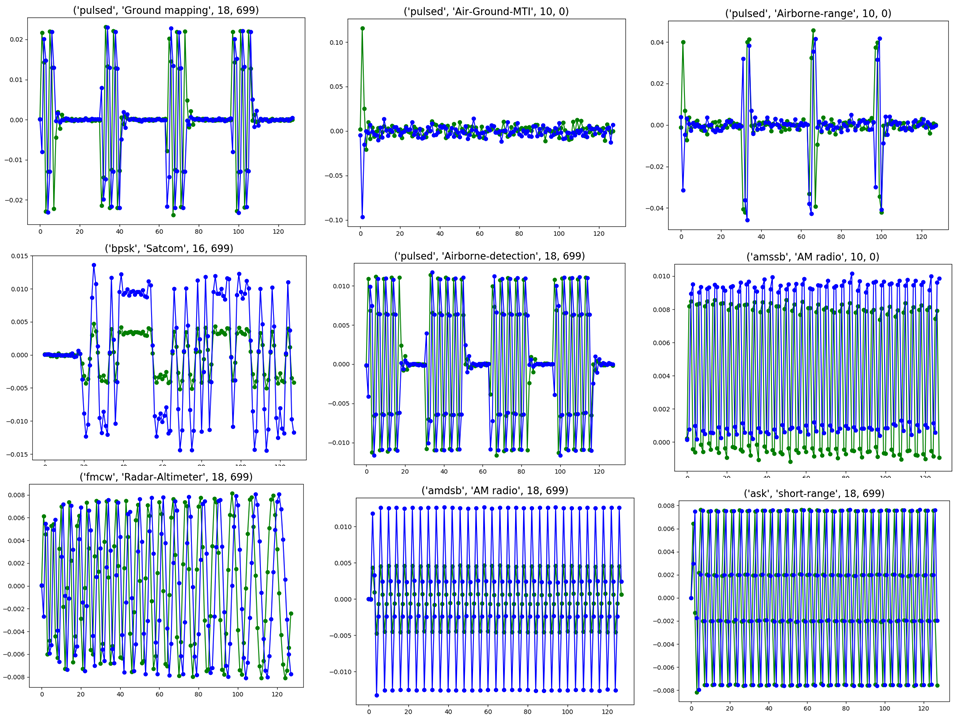} 
\caption{Nine waveforms in RadComDynamic}
\label{fig:9waves}
\end{minipage}
\end{figure*}

\section{DATASET AND SIGNAL PREPROCESSING}
\label{sec:analysis}

\textbf{Dataset and Evaluation Setting:} As ours is the first work in this realm that proposes an MTL architecture for wireless signal recognition, there are no preexisting datasets that could be leveraged with labels for multiple tasks. Hence, we generate our datasets of radar and communication signals in GNU Radio companion \cite{grc} for varying SNRs. We generate two simulated datasets with modeled propagation and/or hardware effects - RadComAWGN and RadComDynamic. RadComAWGN comprises in total of 9 modulation and 11 signal classes as shown in Table \ref{tab:classes}.  
Three signal classes (Bluetooth, IEEE802.11bg, IEEE802.15.4) are extracted from the interference dataset \cite{crawdad}. The remaining waveforms are generated in GNU Radio with additive white Gaussian noise (AWGN) under varying SNR levels (-20 dB to 18 dB in steps of 2 dB) \cite{AjagannathComnet2021}. The SNR levels were applied to the signal by setting the noise amplitude in the GNU Radio blocks (\textit{noise\_source} for RadComAWGN and \textit{dynamic\_channel\_model} for RadComDynamic) as $10^{-SNR/10}$. The RadComDynamic dataset contains all waveforms in RadComAWGN except the 3 waveforms from the interference dataset. The waveforms in the RadComDynamic dataset are subject to propagation effects and hardware uncertainties as shown in Table \ref{tab:dyn}. All 9 distinct waveforms belonging to the RadComDynamic dataset is  shown in Fig \ref{fig:9waves}. We divided each of our dataset into 70\% training, 20\% validation, and 10\% testing sets. The hyper-parameter evaluations were performed with the RadComAWGN dataset. All waveforms of the RadComDynamic were also generated experimentally OTA to derive RadComOta dataset. This is elaborated in section \ref{sec:ota}. To benefit future research in MTL on wireless signals, we make the RadComDynamic and RadComOta datasets publicly available \cite{data}.

\subsection{Wireless Signal Representation}
\label{sec:sig}
The generated 128 sample complex (IQ) signal vector is denoted as $\mathbf{y}^{id}$ where $id$ indicates the key with which it is extracted from the dataset container.
The signals are normalized to unit energy prior to storing them in the dataset to remove any residual artifacts from the simulated environment. Data normalization allows a neural network to learn the optimal parameters quickly thereby improving the convergence properties. The normalized data containing both I and Q samples can be denoted as $\hat{\mathbf{y}}^{id} = \hat{\mathbf{y}}^{id}_I + j\hat{\mathbf{y}}^{id}_Q$. Since neural networks can only deal with real numbers, we will vectorize the complex number as below
$\hat{\mathbf{y}}^{id}$
\begin{equation}
   f\{\hat{\mathbf{y}}^{id}\} = \begin{bmatrix} \hat{\mathbf{y}}^{id}_I\\\hat{\mathbf{y}}^{id}_Q\end{bmatrix} \in \mathbb{R}^{256\times1}
\end{equation}
Mathematically, this can be shown with the relation
\begin{equation}
    f:\mathbb{C}^{128\times1} \longrightarrow \mathbb{R}^{256\times1}
\end{equation}
If the first layer is a convolutional layer, the 256-sample input signal is reshaped to a 2D tensor of size $16\times16$ prior to feeding into the network. In the discussion herein, we fix the task loss weights at $w_{s} = 0.8$ and $w_{m} = 0.2$. The empirical basis for this determination can be referred in our previous work \cite{AJagannath21ICC}. 

\section{NEURAL NETWORK ARCHITECTURE}

How dense should the network be? This is the question we are trying to answer in this section. Resource constrained radio platforms require lightweight neural network models for implementation on general purpose processors (GPPs), field programmable gate arrays (FPGAs) and application-specific integrated circuits (ASICs). For such realistic implementations, dense neural network models for signal characterization such as the resource-heavy AlexNet and GoogLeNet adopted by \cite{AlexGoogleNet_AMC} would seem impractical. Hence, rather than adopting dense computer vision models, we handcraft the MTL architecture to arrive at a lighter model. \emph{Specifically, we are trying to design a small network which is yet big enough to support as many classes corresponding to the tasks.} It is to be noted that the dataset used for the architecture analysis in sections \ref{sec:tw} through \ref{sec:kernel} is the RadComAWGN.

\subsection{Task Weights}
\label{sec:tw}
In this subsection, we will study the effect of task-specific loss weights on the classification accuracy of both tasks. Specifically, the classifier accuracy on both tasks when the signal strength is very low (SNR$=-2$ dB) will be analyzed. Detection of even the weakest power signal corresponds to improved detection sensitivity. 

\begin{figure}[t!]
\centering
\includegraphics[width=0.85\columnwidth]{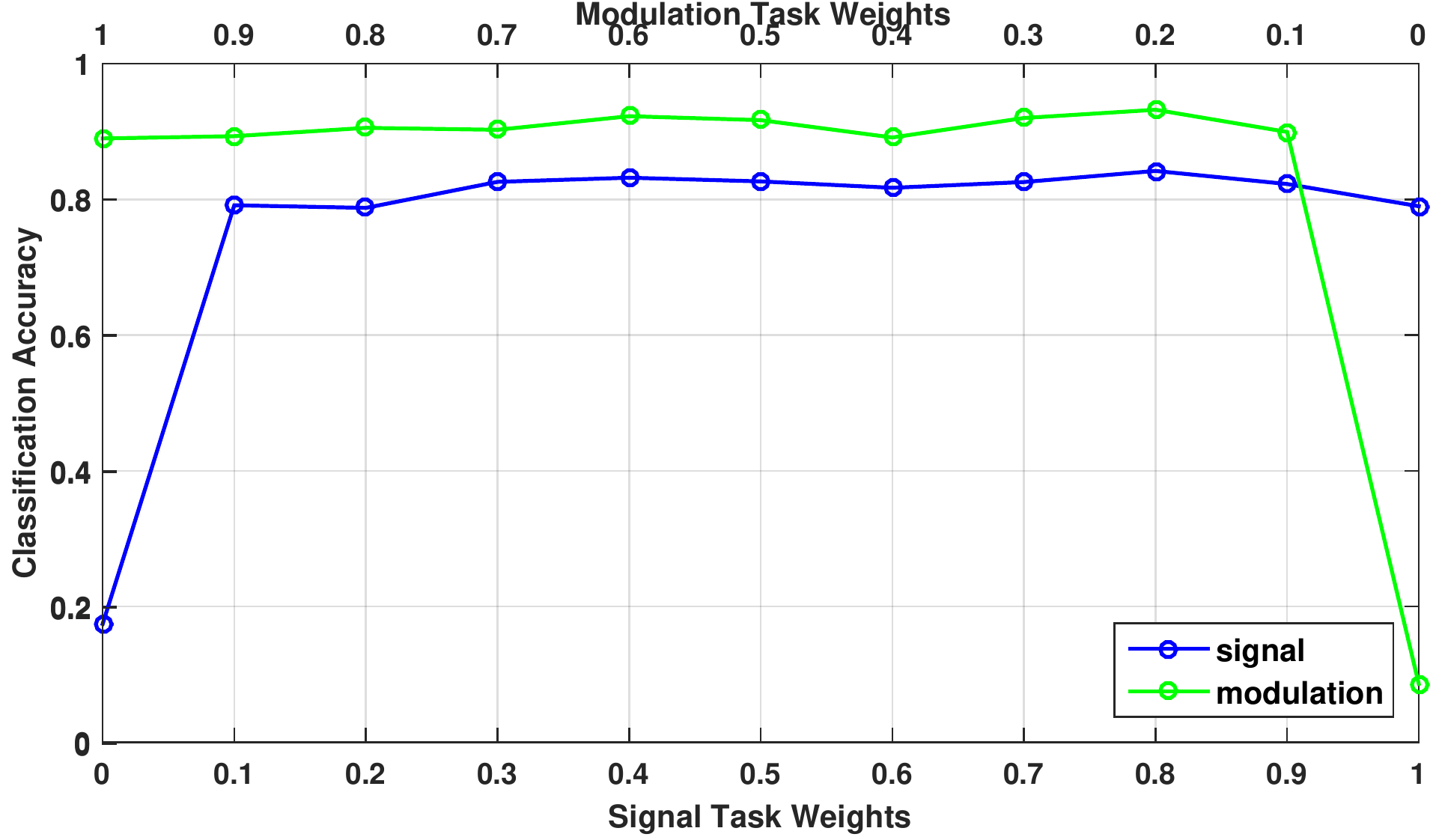} 
\caption{Effect of task loss weight distribution on modulation and signal classification tasks at very low SNR ($-2$ dB)} 
\label{fig:task_weight}  
\end{figure}

Fig. \ref{fig:task_weight} shows the classification accuracy of MTL on both tasks at a very low SNR of $-2$ dB for varying weights. The number of kernels in the shared and task-specific convolutional layers are $8$ and $4$ respectively and the number of neurons in the fully-connected layers of the task-specific branches is $256$. The weight distribution for both tasks are varied from $0$ to $1$  in steps of $0.1$ such that sum of weights is unity. The boundaries of the plot denote classification accuracies when the model was trained on individual tasks, i.e., when weights of either task losses were set to zero. It can be seen that the model performs almost stable across the weighting ($0.1$ to $0.9$ on either task). Although for some optimal weighting of $w_{s} = 0.8$ and $w_{m} = 0.2$, both tasks are performing slightly better than at other task weights. We therefore fix the loss weights for both tasks at $w_{s} = 0.8$ and $w_{m} = 0.2$ for the proposed MTL architecture. 

\subsection{Number of Layers, Neurons}
\label{sec:neurons}

\begin{figure*}
\begin{minipage}[]{0.47 \linewidth}
\centering
\includegraphics[width=0.9\columnwidth]{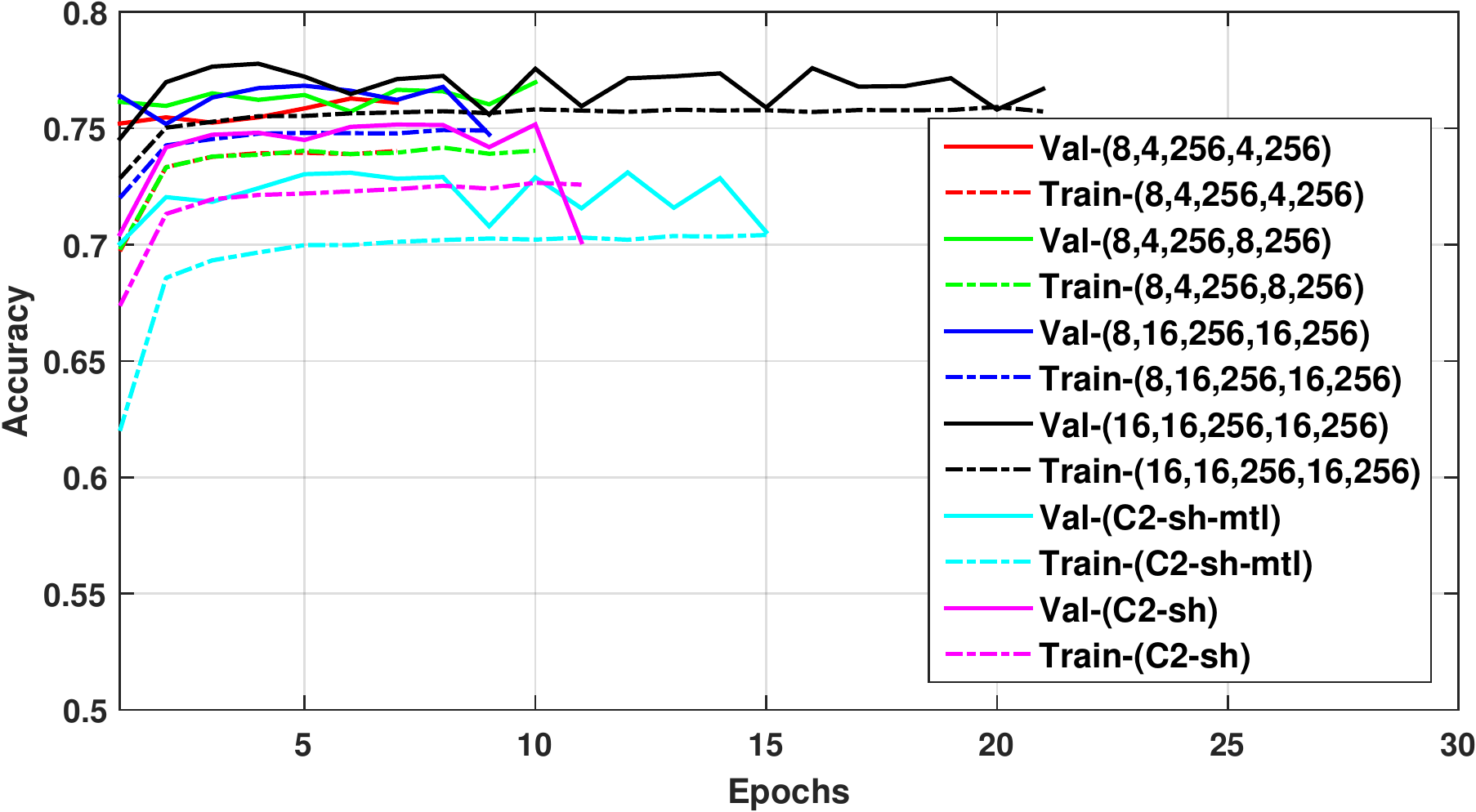} 
\caption{MTL training performance on modulation classification task for varying network density}
\label{fig:mod_nwtrain} 
\end{minipage}\hspace{0.5 cm}
\begin{minipage}[]{0.47 \linewidth}
\centering
 \includegraphics[width=0.9\columnwidth]{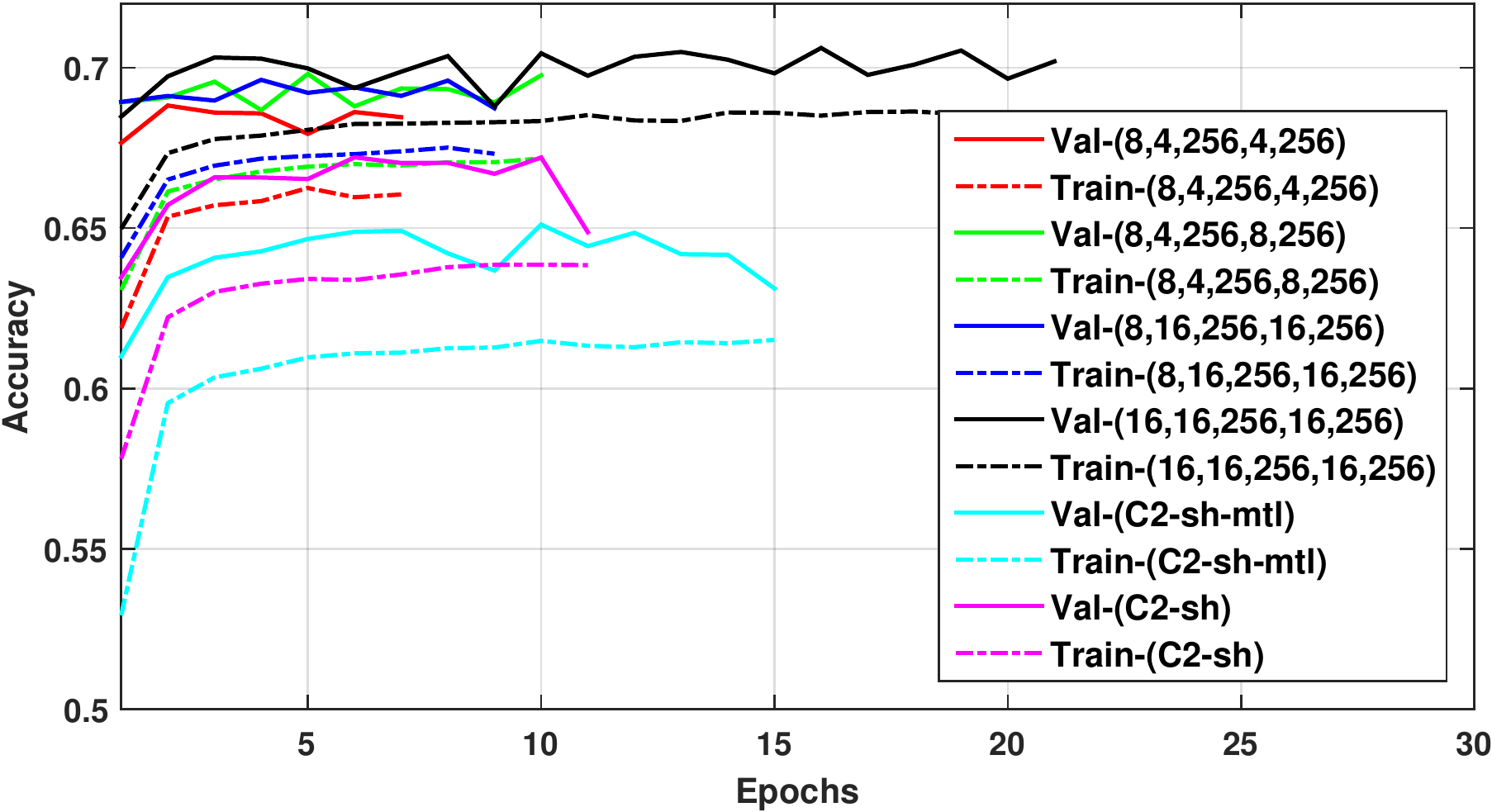} 
\caption{MTL training performance on signal classification task for varying network density}
\label{fig:sig_nwtrain}
\end{minipage}

\begin{minipage}[]{0.47 \linewidth}
\centering
\includegraphics[width=0.9\columnwidth]{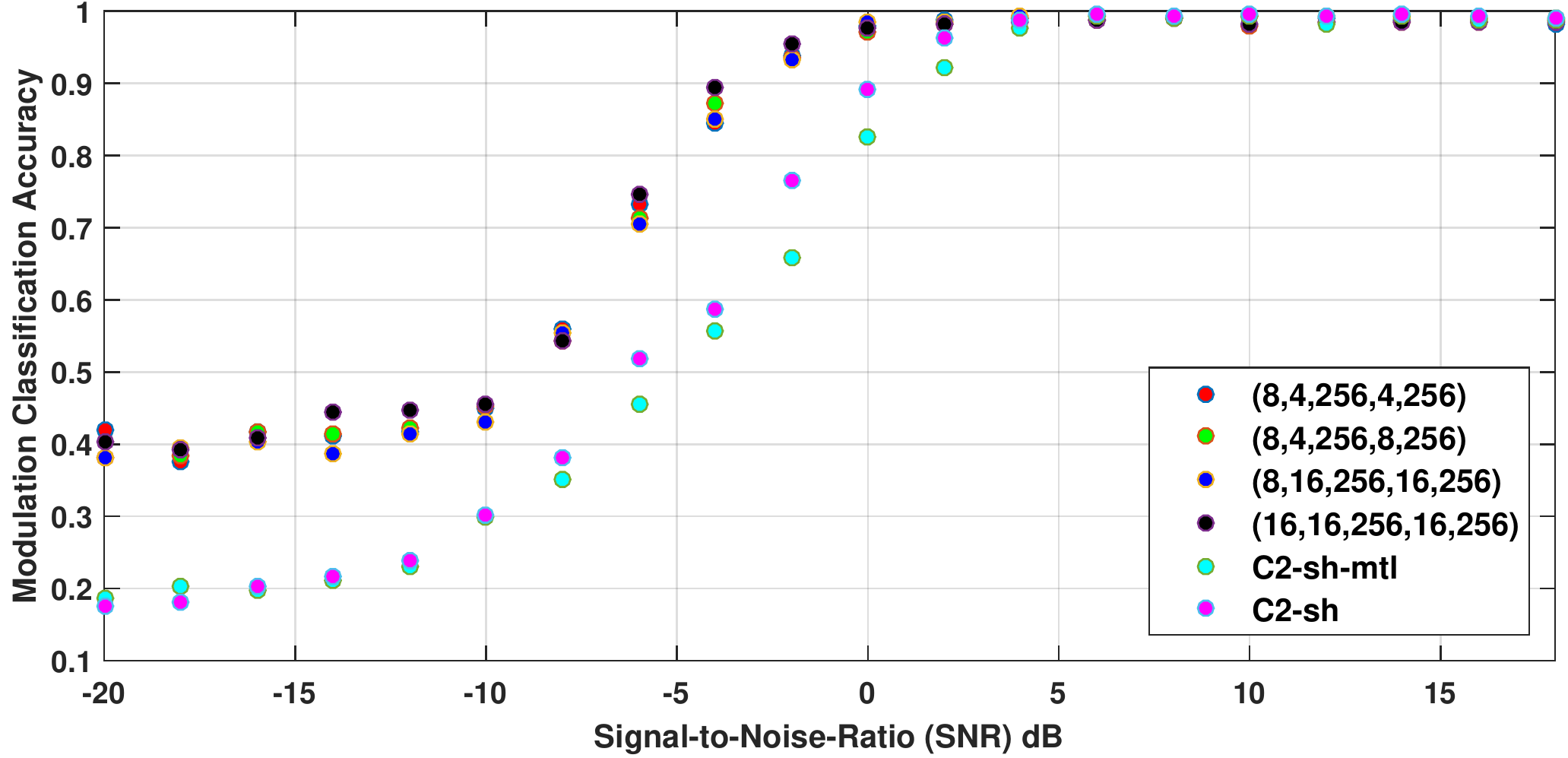} 
\caption{MTL classification performance under varying noise levels on modulation task for varying network density}
\label{fig:mod_nw} 
\end{minipage}\hspace{0.5 cm}
\begin{minipage}[]{0.47 \linewidth}
\centering
\includegraphics[width=0.9\columnwidth]{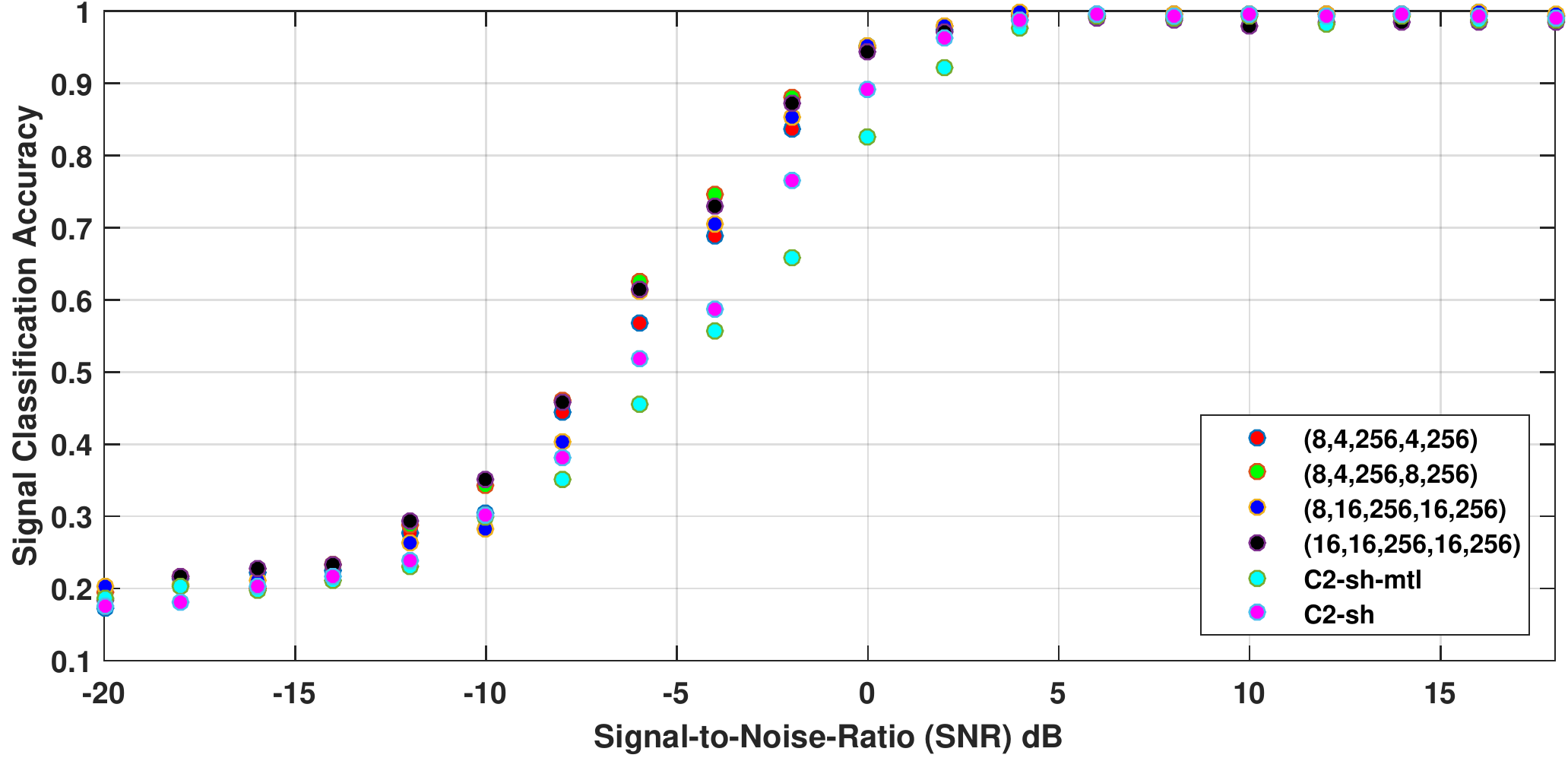} 
\caption{MTL classification performance under varying noise levels on signal task for varying network density}
\label{fig:sig_nw}  
\end{minipage}
\end{figure*}

We will vary the number of neurons in the MTL model introduced in Fig. \ref{fig:mtl} and analyze the effect of introducing additional layers in the shared as well as task-specific branches. 

The legends in the figures (Fig. \ref{fig:mod_nwtrain} - Fig. \ref{fig:sig_nw}) represent the varying number of neurons as well as layers in the network. The notation $(C_{sh},C_{m},F_{m},C_{s},F_{s})$ implies neuron distribution with $C_{sh},C_{m},C_{s}$ representing the number of filters in the convolutional layer of shared, modulation, and signal branches and $F_{m}, F_{s}$ denote the number of neurons in the fully-connected layers in the modulation and signal branches. The additional layer inclusion notations are $C2-sh$ and $C2-sh-tasks$. The notation $C2-sh$ denotes the MTL architecture with two convolutional layers each followed by a max-pooling layer in the shared trunk. The number of filters in the convolutional layers of the shared trunk is 8. Finally, $C2-sh-tasks$ denote the MTL model with shared branch composition the same as $C2-sh$ but with two sequential convolutional layers in the task-specific branches. The number of filters in the convolutional layers of both task-specific branches is 4. The number of neurons in the fully-connected layers of task-specific branches is 256 for both $C2-sh$ and $C2-sh-tasks$.

Fig. \ref{fig:mod_nwtrain} and Fig. \ref{fig:sig_nwtrain} show the training performance of the MTL model with respect to the two tasks. The training plots demonstrate that increasing the network density slows the training speed of the model. This is intuitive as the network parameters increase training time increases. The fastest network training time is achieved with the model configuration of $(8,4,256,4,256)$ which is the lightest of all configurations. Fig. \ref{fig:mod_nw} and Fig. \ref{fig:sig_nw} demonstrate the classification accuracy on both tasks for varying network density under increasing SNR levels (decreasing noise power). It can be seen that the additional layers in the shared ($C2-sh$) and shared as well as task-specific branches ($C2-sh-tasks$) does not improve the classification accuracy but rather results in significantly poor modulation and signal classification accuracy. Further, the MTL model does not seem to benefit from the remaining dense configurations. Hence, the MTL model will use the lighter configuration of $(8,4,256,4,256)$ that yields better learning efficiency and prediction accuracy.

In a nutshell, we empirically evaluated that the introduction of additional layers in the shared and task-specific branches does not improve the classification accuracy but rather results in significantly poorer modulation and signal classification accuracy. Out of the various evaluated configurations, we determined 8 and 4 convolutional kernels with 1 convolutional layer each in the shared and task-specific branches respectively and 256 fully connected neurons in the single fully connected layer of both task branches notated in \cite{AJagannath21ICC} as $(8,4,256,4,256)$ yields better learning efficiency and prediction accuracy.

\subsection{Convolutional Kernel Size}
\label{sec:kernel}
The kernel sizes of convolutional filters influence the memory and compute requirements, training, and classification performances of the model. The number of parameters and computations scale with the kernel size as well as the number of kernels. To understand this better, let us consider an example where the input to the convolutional layer has dimensions $16\times16\times1$ (considered in this paper). Let us suppose the convolutional kernel has dimensions $7\times7\times1$ with 8 kernels of input padding size $=0$ and stride$=1$. This yields total number of learnable parameters to be $7\times7\times1\times8 = 392$ and total number of computations as $10\times10\times8\times7\times7=39200$. With the same number of kernels, stride, and padding size, a kernel size of $3\times3\times1$ will result in $72$ parameters and $14112$ computations. This tells us a kernel size of $3\times3$ result in $\sim5.4\times$ lesser memory and $\sim2.7\times$ fewer computations.

In this evaluation, the convolutional kernel sizes in the shared and task-specific branches are increased from $3\times3$ to $7\times7$. Recall here (from Section \ref{sec:neurons}) that we use 8 kernels in the shared branch and 4 in both task-specific branches. 
\begin{figure}[h!]
\centering
\includegraphics[width=0.9\columnwidth]{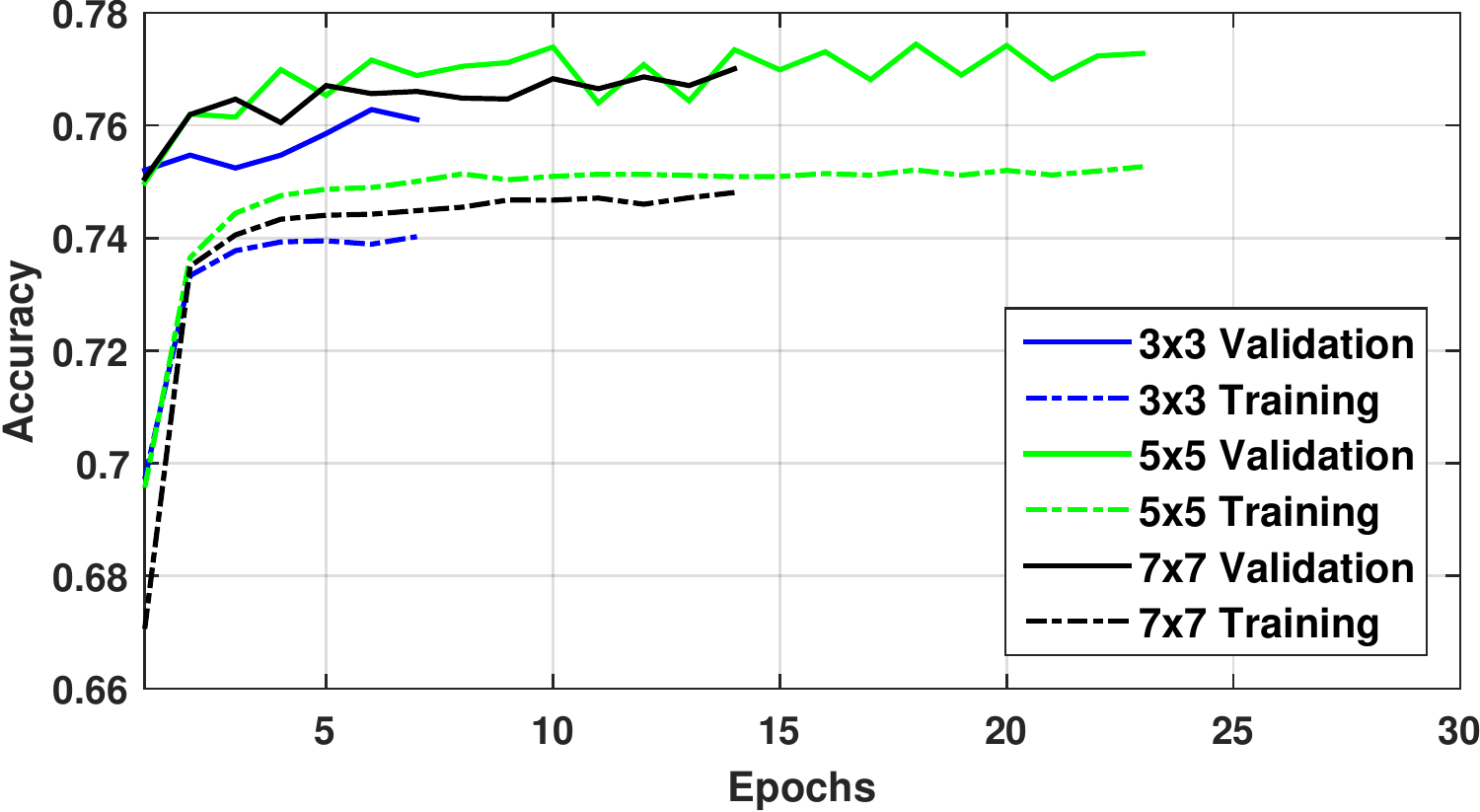} 
\caption{MTL training performance on modulation classification task for varying kernel size}
\label{fig5}
\end{figure}

\begin{figure}[h!]
\centering
\includegraphics[width=0.9\columnwidth]{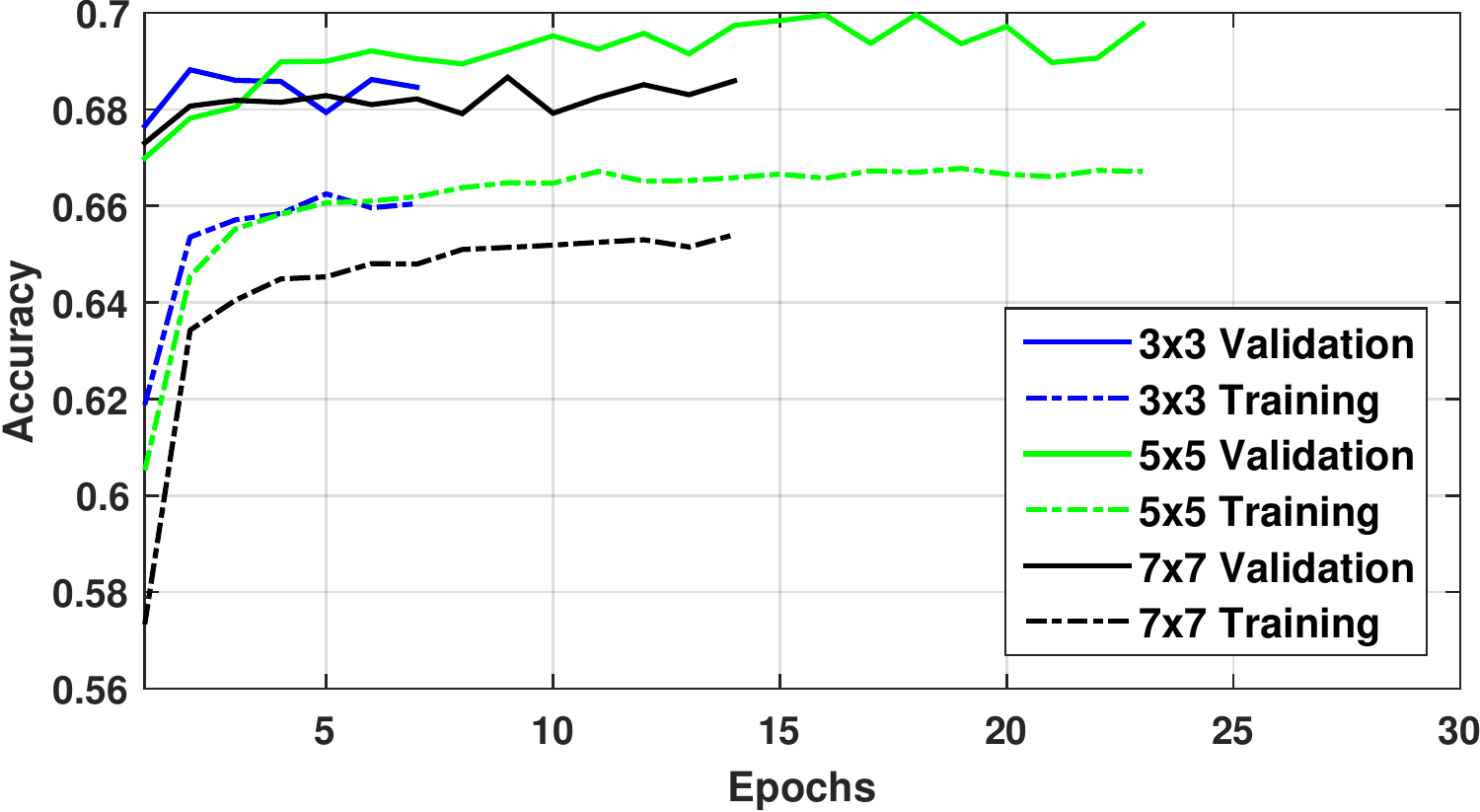} 
\caption{MTL training performance on signal classification task for varying kernel size}
\label{fig6}
\end{figure}

\begin{figure}[h!]
\centering
\includegraphics[width=0.9\columnwidth]{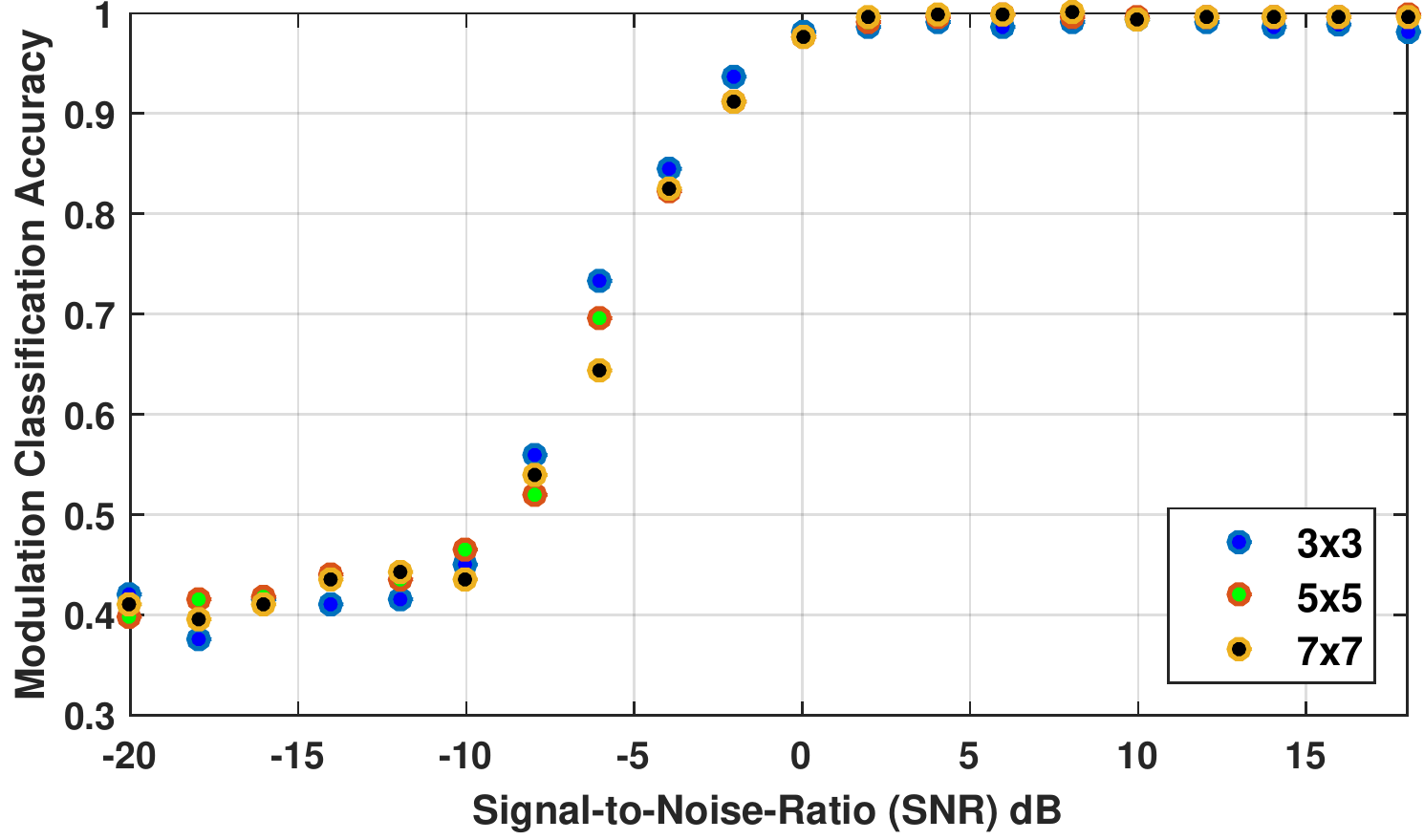} 
\caption{MTL classification performance under varying noise levels on modulation task for varying kernel size}
\label{fig7}
\end{figure}

\begin{figure}[h!]
\centering
\includegraphics[width=0.9\columnwidth]{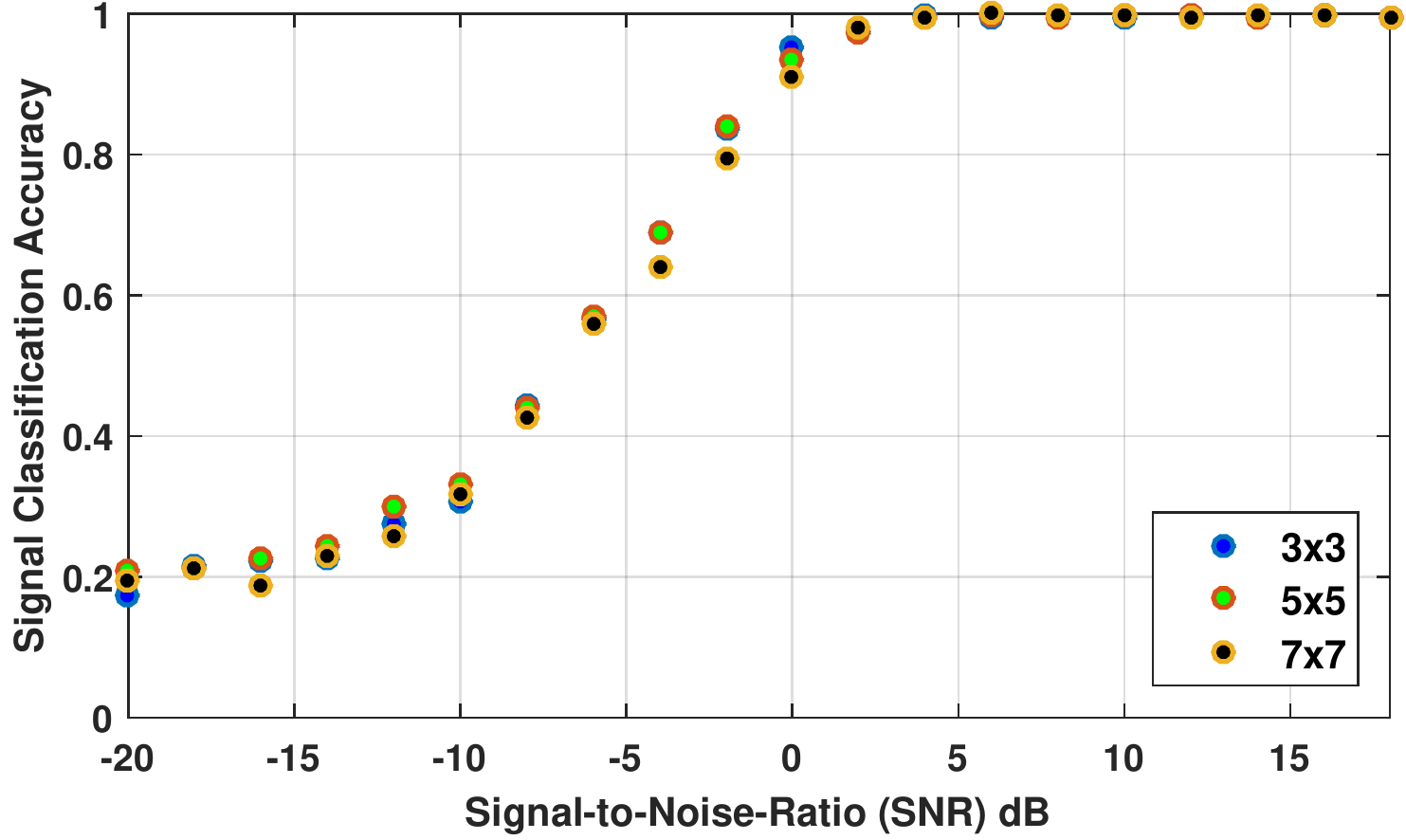} 
\caption{MTL classification performance under varying noise levels on signal task for varying kernel size}
\label{fig8}
\end{figure}
Fig. \ref{fig5} and Fig. \ref{fig6} depict the training speed of the model with varying convolutional kernel sizes under varying noise levels. The plots demonstrate faster training of the model with the smallest kernel size of $3\times3$. From the classification accuracy point of view, Fig. \ref{fig7} and Fig. \ref{fig8} demonstrate no significant benefit from employing larger convolutional kernel sizes. Therefore, for the considered number of classes corresponding to the tasks, we choose a memory and computation efficient design choice of $3\times3\times1$ kernel size in all branches of the MTL. The final deduced MTL architecture following these in-depth evaluations are shown in Fig. \ref{fig:mtl_arch}. Please note here that the "$?$" in the architecture indicates the batch size for the model training.

\subsection{Feature Extraction}
\label{fe}

\begin{figure*}[t!]
\centering \vspace{-.2 cm}
\includegraphics[width=1.4\columnwidth]{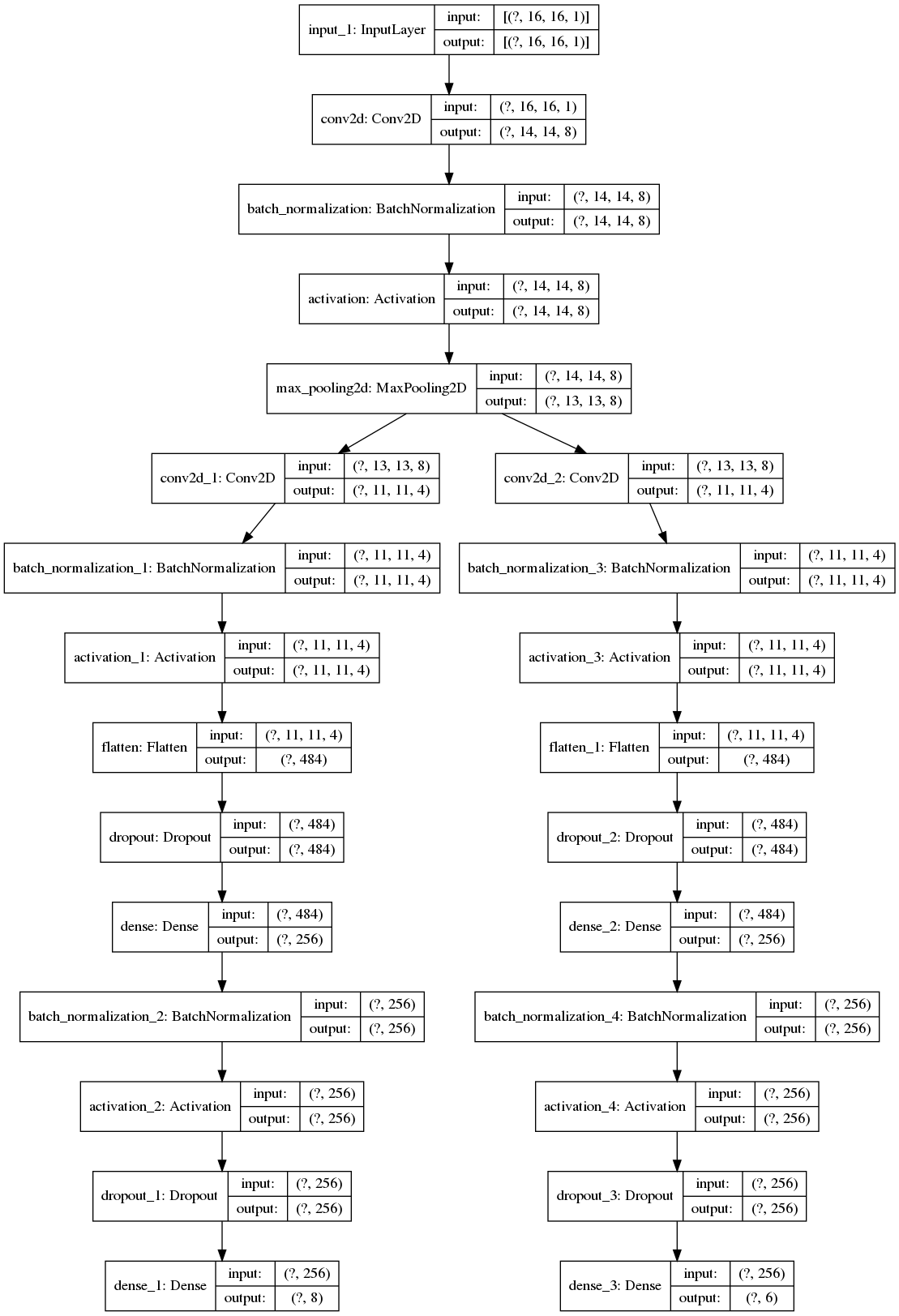} 
\caption{Deduced MTL architecture}
\label{fig:mtl_arch}  
\end{figure*}

As stated in section \ref{sec:sig}, the IQ vector $\mathbf{y}^{id}$ is reshaped into a squared tensor of dimension $16\times16\times1$ in order to enable the convolutional layer (conv2d denoted in Fig. \ref{fig:mtl_arch}) to perform the spatial feature extraction. Consider an input feature map $F_i\in \mathbb{R}^{N_i\times H_i\times W_i}$ fed to a convolutional layer $i$ where $N_i,\; H_i\;, W_i$ are the number of input channels, height, and width of the input feature map, respectively. Let the layer $i$ contain $N_{i+1}$ convolutional kernels of dimensions $h_k\times w_k\times W_i$ corresponding to a total number of learnable parameters of $N_ih_kw_kW_i$. The convolutional layer maps the input feature maps to output tensor $F_{i+1}\in\mathbb{R}^{N_{i+1}\times H_{i+1}\times W_{i+1}}$ which serves as input for the next convolutional layer $i+1$ by the following transformation. 
\begin{equation}
    F_{i+1}(m_p,n_q) = \sum_{p=1}^{h_k}\sum_{q=1}^{w_k}\sum_{r=1}^{W_i}K_i(p,q,r)F_i(m,n)
\end{equation}
where the spatial location of the output are $m_p=m-p+1$ and $n_q=n-q+1$ considering a unit stride without zero-padding.
In other words, each convolutional kernel in layer $i$ of size $h_k\times w_k\times W_i$ generates one feature map. The total number of floating point operations (FLOPs) of layer $i$ is $N_{i+1}H_{i+1}W_{i+1}h_kw_kW_i$. We can see that the shared branch only has a single convolution layer - conv2d - followed by a pooling layer - max\_pooling2d. The conv2d layer has 8 kernels of size $3\times3\times1$ generating a feature map of size $14\times14\times8$ which is sub-sampled by the pooling layer of size $2\times2$ to $13\times13\times8$. The FLOPs of the conv2d layer can be obtained as $14\times14\times8\times3\times3\times1 = 14.112$k. Additionally, note here that we adopt batch normalization prior to ReLU activation to reduce the variance across samples in a batch \cite{BN}. We have empirically determined faster convergence by adopting batch normalization.

The feature map of size $13\times13\times8$ from the shared branch is fed into the two task branches that possess a convolution layer and a fully connected layer each. The convolution layer in these task branches has 4 kernels of size $3\times3\times1$ transforming the $13\times13\times8$ feature map into $11\times11\times4$ feature tensor. A flattening (vectorize) operation is performed as indicated by Flatten in the Fig. \ref{fig:mtl_arch} to produce a 1D vector from the 3D tensor which is fed to the fully connected layer (Dense) of 256 neurons. We adopt dropout as regularization to stabilize the convergence of the model. The final output layers of the two tasks are denoted by dense\_1 and dense\_3 which performs softmax classification.

\begin{figure*}[t!]
\centering
\includegraphics[width=1.7\columnwidth]{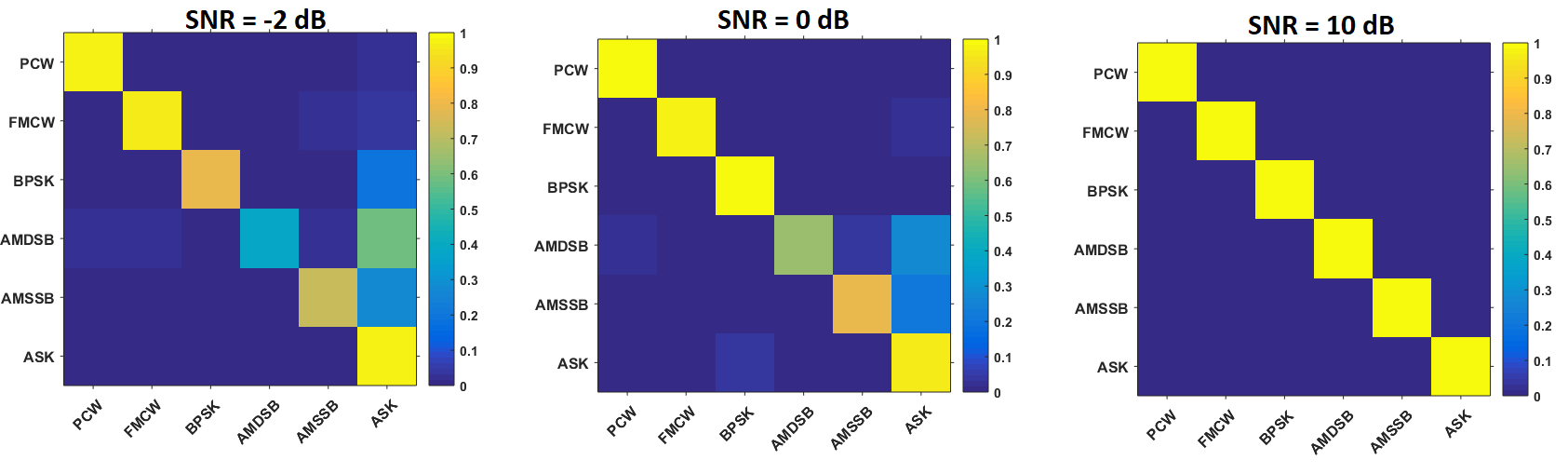} 
\caption{Fine-tuned MTL modulation classification: Confusion matrices at -2 dB, 0 dB, and 10 dB with challenging RadComDynamic dataset.}
\label{fig:fig14}
\end{figure*}

\begin{figure*}[h!]
\centering
\includegraphics[width=2\columnwidth]{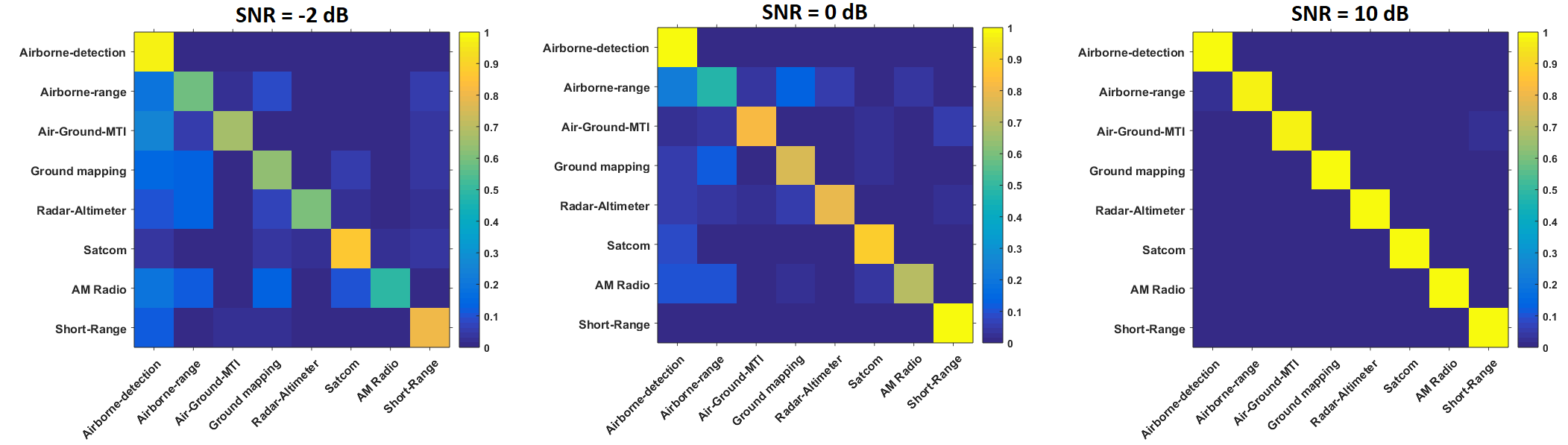} 
\caption{Fine-tuned MTL signal classification: Confusion matrices at -2 dB, 0 dB, and 10 dB with RadComDynamic dataset.}
\label{fig:fig15}
\end{figure*}

\begin{figure}[h!]
\centering
\includegraphics[width=0.9\columnwidth]{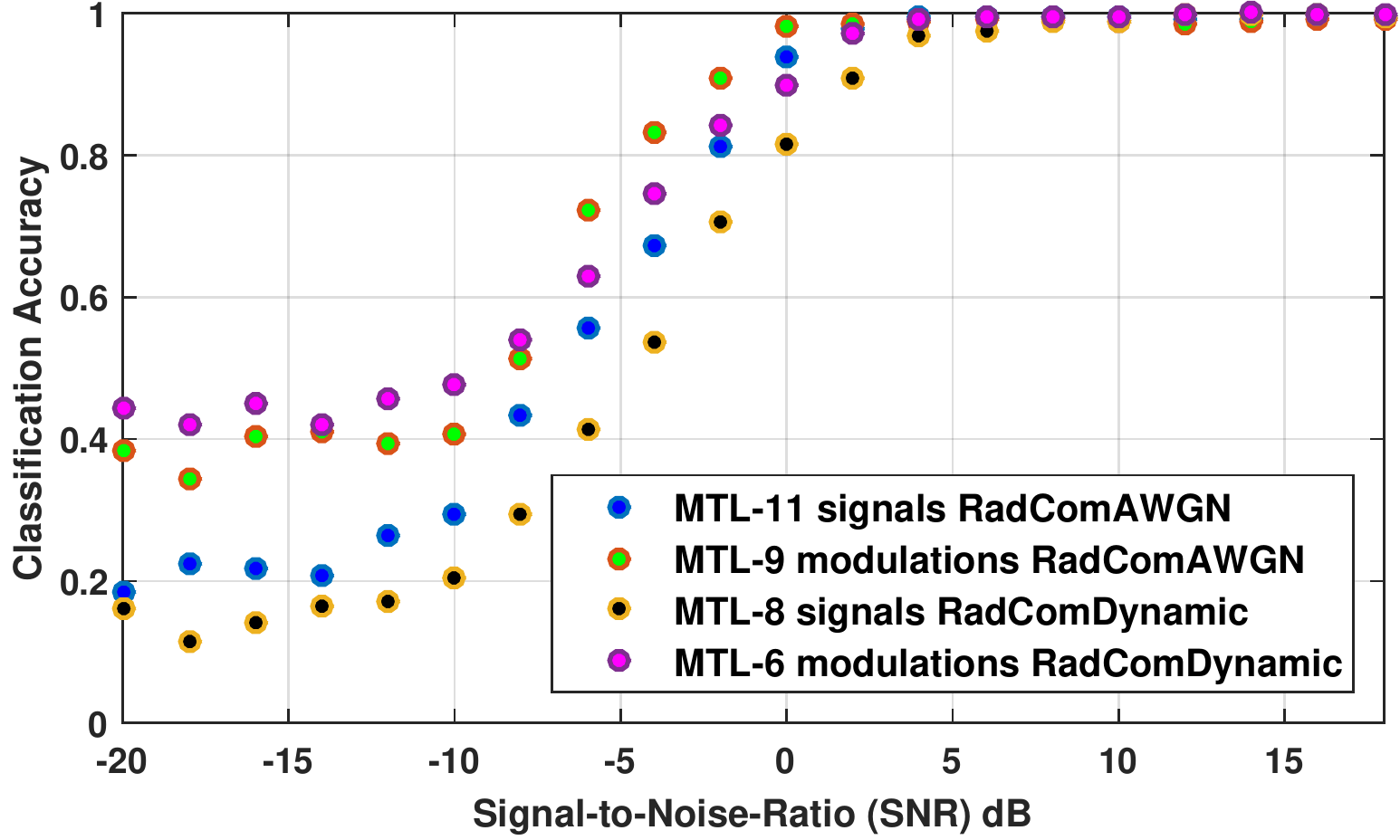} 
\caption{Fine-tuned MTL classification under varying SNR with RadComDynamic dataset.}
\label{fig:fig16}
\end{figure}
\begin{table*}[ht!]
\footnotesize
    \centering
    \def\arraystretch{1.4}%
    \caption{Summary of related works \label{tab:para}}
    \begin{tcolorbox}[tab2,tabularx={|p{3.1cm}||p{3.7cm}||p{3.1cm}||p{2.6cm}||p{3cm}}]
\hline
\textbf{Model}       & \textbf{Modulation Classification Acc.}       & \textbf{Signal  Classification Acc.}       &\textbf{Number of Classes}       &\textbf{Waveform Domain}       \\ \hline\hline
\multicolumn{5}{|c}{\textbf{Modulation and Signal classification (this work) - Multi-task}} \\ \hline\hline
Proposed MTL Model         &97.87\% at 0 dB, \newline 99.53\% at 10 dB                &92.3\% at 0 dB,\newline 99.53\% at 10 dB                &9 modulation,\newline 11 signal classes              &Radar and \newline Communication                   \\ \hline\hline
\multicolumn{5}{|c}{\textbf{Modulation classification only methods - Single Task}}                                  \\ \hline\hline
Oshea and West \cite{vtcnn} & $\sim 70\%-75\%$ at 18 dB & - & 11 modulation & Communication\\\hline
Wang, Liu,Yang, and Gui \cite{drcnn} & $> 95\%$ at $>$2 dB & - & 8 modulation & Communication\\\hline
Peng et al. 2019 \cite{AlexGoogleNet_AMC}        & below 80\% at 0 dB               & -              & 8 modulation             &Communication                    \\ \hline
Jagannath et al. 2018 \cite{Jagannath18ICC}        & 98\% above ~25 dB               & -              & 7 modulation             &Communication                    \\ \hline
O'Shea et al. 2018 \cite{oshea2018}        & 95.6\% at 10 dB               & -              & 24 modulation              &Communication                    \\ \hline
Mossad et al. 2019 \cite{mtlmod}        & 86.97\% at 18 dB               & -              & 10 modulation              &Communication                    \\ \hline
Hermawan et al. 2020 \cite{ICAMCNet}        &$\sim$80\% at 0 dB, \newline 83.4\% at 18 dB              & -              & 11 modulation              &Communication                    \\ \hline
Wang et al. 2017 \cite{radar_recog}        &100\% at 0 dB              & -              & 7 modulation              &Radar                    \\ \hline
Xu et al. \cite{XuLiWang2019}        &$\sim$100\% at 0 dB              & -              & 5 modulation             &Communication    \\ \hline
Li et al. 2018 \cite{vhf_amc}        &95\% above 2 dB              & -              & 7 modulation              &Communication    \\ \hline\hline
\multicolumn{5}{|c}{\textbf{Signal classification only methods - Single Task}}                                   \\ \hline\hline
Bitar et al. 2017 \cite{wirelesstech}        &-              &91\% at 15-25 dB, \newline 93\% at 30 dB              & 7 signal classes              &Communication    \\ \hline
Schmidt et al. 2017 \cite{wirelessInterference}  &-              &95\% at -5 dB             & 15 signal classes              &Communication    \\ \hline
\end{tcolorbox}{}
\end{table*}
\subsection{Fine-tuned Model Performance}

In this section, we evaluate the performance of the fine-tuned MTL model on both the datasets (RadComAWGN and RadComDynamic) under varying SNR. The objective of these experiments are to evalaute MTL model on waveforms impaired by AWGN alone and waveforms impacted by \textit{realistic propagation and radio hardware impairments.} Overall, the MTL exhibits a 99.53\% modulation classification accuracy on RadComAWGN and 97.58\% on RadComDynamic dataset at 2 dB. The signal classification accuracy of MTL at 2 dB yielded 97.07\% and 90.79\% on RadComAWGN and RadComDynamic datasets respectively. We show that the proposed MTL model yields above 90\% accuracy at SNRs above 2 dB for both tasks with RadComAWGN waveform (noise impaired) and RadComDynamic waveforms (propagation and hardware impaired). The confusion matrices of the signal and modulation classes at $\{-2, 0, 10\}$ dB on the challenging RadComDynamic dataset are depicted in Fig. \ref{fig:fig14} and Fig. \ref{fig:fig15} respectively. It can be deduced that at -2 dB three of the modulation classes have above 95\% accuracy while the others except AMDSB have above 70\% accuracy. As SNR increases to 0 dB, the accuracy of these classes further improves, with four modulation classes at $>$98\% accuracy and the lowest accuracy being 65\%. The signal classification accuracy on the other hand had several misclassifications although at lower rates when the SNR is -2 dB indicating the complexity of the signal classification task in contrast to the modulation classification task. The misclassification drops as the SNR increases. Further, the top-1 classification accuracy of both signal and modulation classification on both datasets under varying SNRs are shown in Fig. \ref{fig:fig16}. These experiments establish the learning and classification capability of the novel lightweight multi-task model on extremely impaired RF signals under varying SNR.

As discussed earlier, the proposed MTL framework is the first method that accomplishes both tasks with a single model. Since multi-task RF datasets and/or architectures that could be leveraged to make one-on-one comparison is not present in literature, in Table \ref{tab:para}, we show that our model outperforms most single task classifiers in either task. To be consistent with most of the other datasets, the classification accuracy of the proposed MTL model in the table are with the RadComAWGN noise impaired waveforms.  The single task modulation classifier proposed in \cite{radar_recog} which achieves a 100\% accuracy at 0 dB is with fewer classes and utilizing handcrafted input features. Additionally, raw IQ samples allow the model to capture hidden representations and allows additional waveform inclusion with ease without requiring significant model retraining. The model in \cite{XuLiWang2019} achieves a $100\%$ modulation classification accuracy at 0 dB. However, it only classifies only 5 waveforms that are generated with the same frequency and bandwidth while requiring a denser CNN architecture. In contrast, even our noise impaired waveforms are generated for varying carrier frequencies and bandwidth which is more typical of the realistic setting. Overall, the proposed lightweight model has provided reliable performance over several varying scenarios outperforming most state-of-the-art single-task techniques. 

\section{Over-the-air Evaluation}
\label{sec:ota}
\subsection{Over-the-air Data Collection}

\begin{figure}[t]
    \centering
\includegraphics[width=0.9\columnwidth]{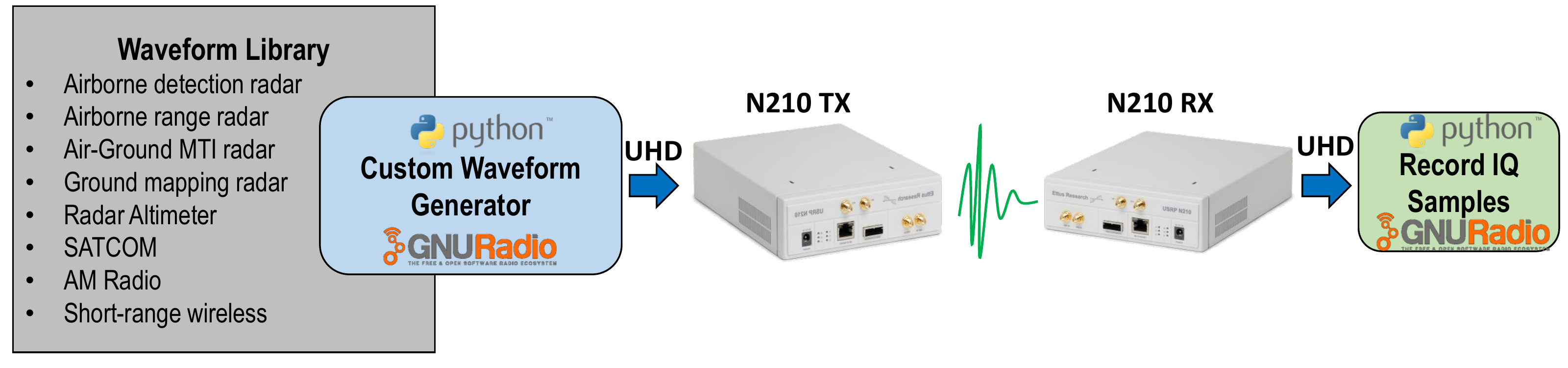} 
\caption{Experimental data collection setup using SDR.}
\label{fig:sdr}    
\end{figure}

We evaluate the performance of the above designed MTL model under an indoor OTA settings - hardware and propagation effects. We use GNURadio to perform the transmission and IQ sample reception using N210 one of the USRP family of SDRs by leveraging the Universal Hardware Driver (UHD) Source and File Sink blocks of GNURadio. The radios use VERT2450 antennas for transmission and reception. The receiver samples the incoming IQ samples at 10 MS/s at a center frequency of 2.45 GHz. Notice, here that we carry out the sample collection under a \emph{in-the-wild} indoor laboratory set up with several other interferences especially from a cluster of WiFi and Bluetooth devices in the vicinity. The receiver gain is set to a constant 30 dB while the transmission gain is varied from 0 to 32 dB in steps of 4 dB for each waveform. The minimum supported transmit gain of the N210 radio is 0 dB. We note here that we do not use the XML/YAML-based GNURadio Companion flowgraph for the data collection setup in Fig. \ref{fig:sdr}, rather we use the Python library of GNURadio to avail certain signal processing blocks. Since the waveforms which include radar and communication are not readily available, these were custom written in Python to create the waveform library and interfaced with USRP SDRs over the UHD API.

\begin{figure}[t]
    \centering
\includegraphics[width=0.8\columnwidth]{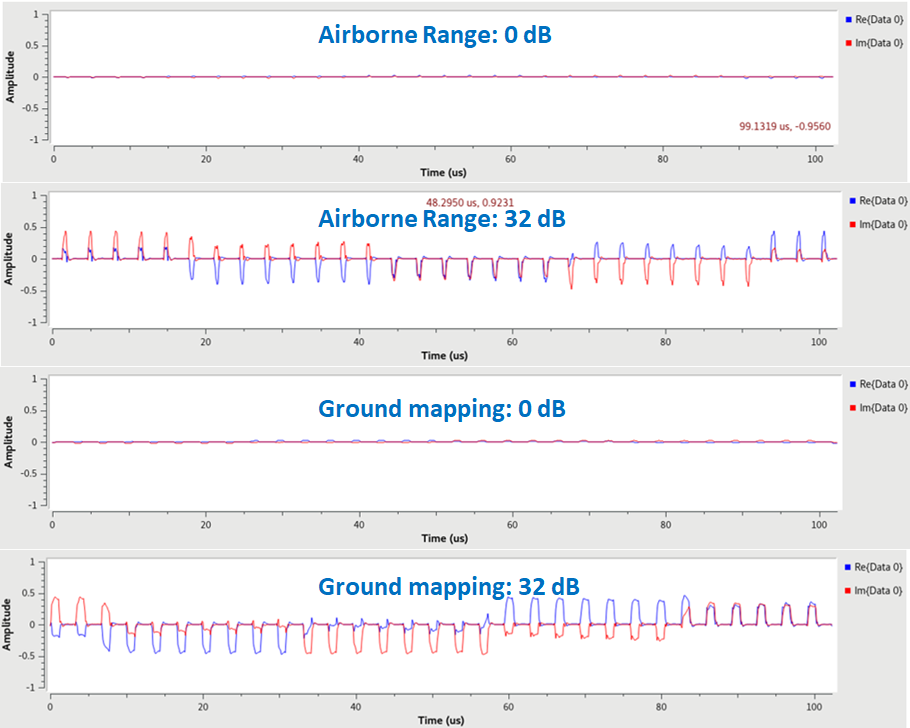} 
\caption{Demonstrating signal amplitudes at different radio transmit gains of Airborne Range and Ground Mapping radar waveforms as seen on the receiving N210.}
\label{fig:amp}    
\end{figure}

We collect all the waveforms belonging to the RadComDynamic dataset wirelessly. For each capture at a certain transmission gain setting, we collect 1.28 million samples. The transmission gain is varied to collect samples under varying signal strengths. In order to benefit future research for practitioners in this realm, we make this experimental OTA collection (RadComOta) accessible for public use \cite{data}. Due to the nature of collection in the presence of other unavoidable interferences as in a real-world setting, this dataset is challenging and more relevant (w.r.t realistic applications) compared to our previous open source dataset - RadComDynamic. The dataset has over 8 Million IQ samples amounting to the six modulation and eight signal classes at transmit gains 0 to 32 dB in steps of 4 dB where each transmit gain setting for a specific waveform has 7k examples each of 128 IQ samples. To provide intuition as to how low the signal power is at 0 dB, we show the signal amplitudes of Airborne Range and Ground mapping radars and compare it to 32 dB transmission in Fig. \ref{fig:amp}.

\subsection{Benchmark Architectures}

In order to evaluate and benchmark the proposed MTL model, we adopt two reference architectures from the literature - VTCNN \cite{vtcnn} and DRCNN \cite{drcnn}. These architectures ingest 128 complex IQ samples as in the proposed MTL, thereby serving as a good means to benchmark the performance using the same dataset. Recall here that there are no other existing MTL architectures for the wireless signal recognition application to date. Hence, we modify the output layer of the reference architectures to accommodate multiple tasks classification.

\begin{figure*}[t!]
\centering 
\includegraphics[width=1.6\columnwidth]{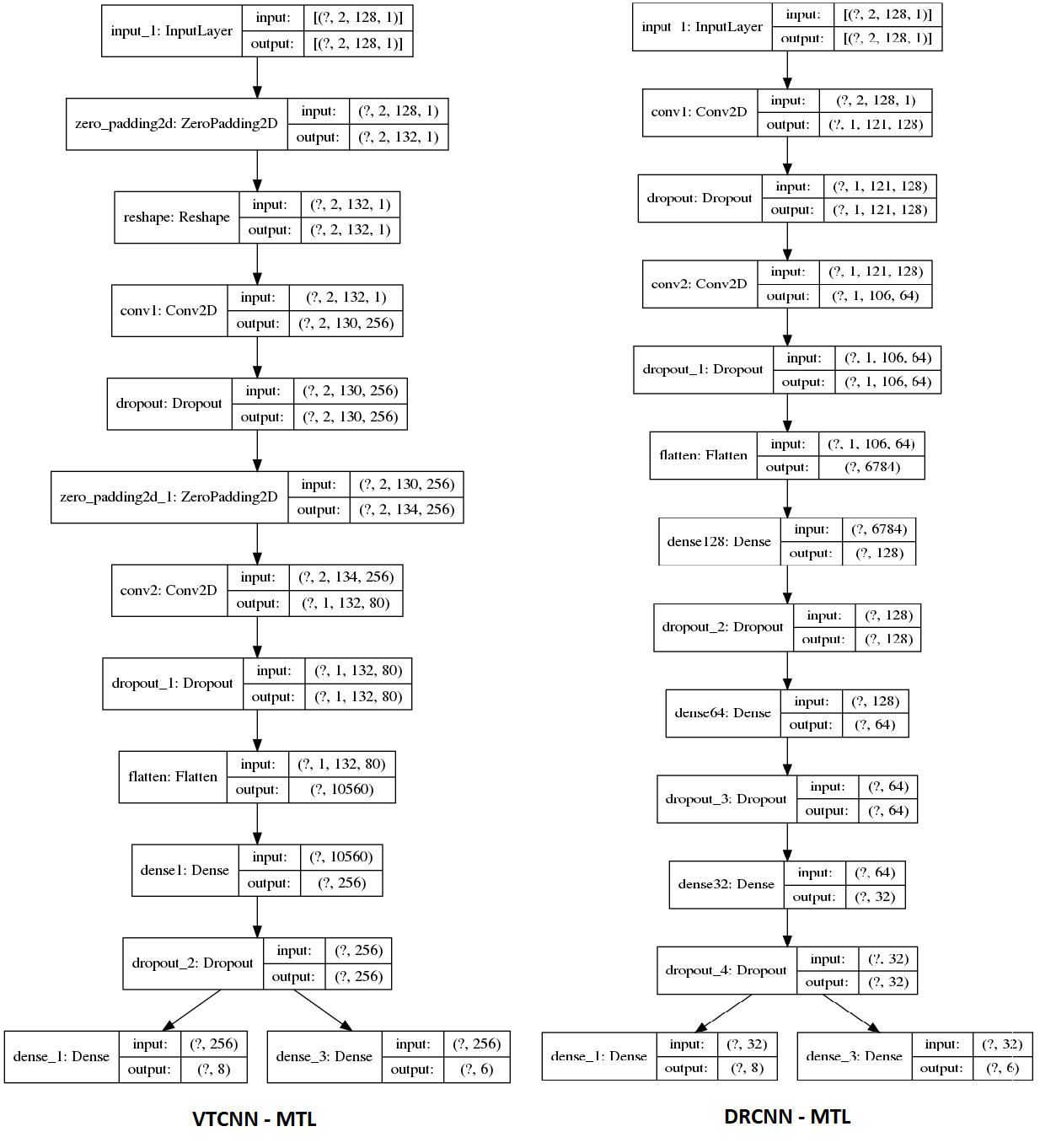} 
\caption{Benchmark architectures}
\label{fig:b_arch}  
\end{figure*}

\subsection{MTL Prediction with Over-the-air Data}

In order to measure the performance with OTA captured dataset, we partition the dataset similar to the synthetic case into 70\% training, 20\% validation, and 10\% test. The fine-tuned MTL architecture is then trained on the OTA dataset. We considered data augmentation with Gaussian noise layer such that Gaussian noise is added to each training sample, hence improving the overall performance by preventing the network from learning irrelevant features. We would like to state here that \emph{any neural network is only as good as the data being fed}. A neural network tends to overfit when attempting to learn high frequency features (patterns that occur more frequently). Zero mean Gaussian noise has data points at all frequencies and thereby dampens the high frequency features. In Keras, we achieve the noise induced data augmentation by using \emph{GaussianNoise} layer with configurable standard deviation (0.1 in this evaluation). This layer is inserted as the last layer in the shared branch. We note here that this layer acts as a regularization layer and is only active during training.
\begin{table}[h!]
    \centering
    \caption{Model accuracy at lowest radio transmit gain} \label{tab:bacc}
    \begin{tcolorbox}[tab2,tabularx={|p{2.5 cm}||p{2.8 cm}||p{2.8 cm}}]
      \textbf{Model}   & \textbf{Modulation @ 0 dB} &\textbf{Signal @ 0 dB}\\ \hline\hline
      
      Proposed MTL model & \textbf{82.4}\% &\textbf{90.1}\% \\ \hline
      VTCNN-MTL &80.14\% &87.65\% \\ \hline
      DRCNN-MTL &77.5\% &88.2\% \\ \hline
    \end{tcolorbox}{}
\end{table}

\begin{figure}
\centering
\includegraphics[width=0.99\columnwidth]{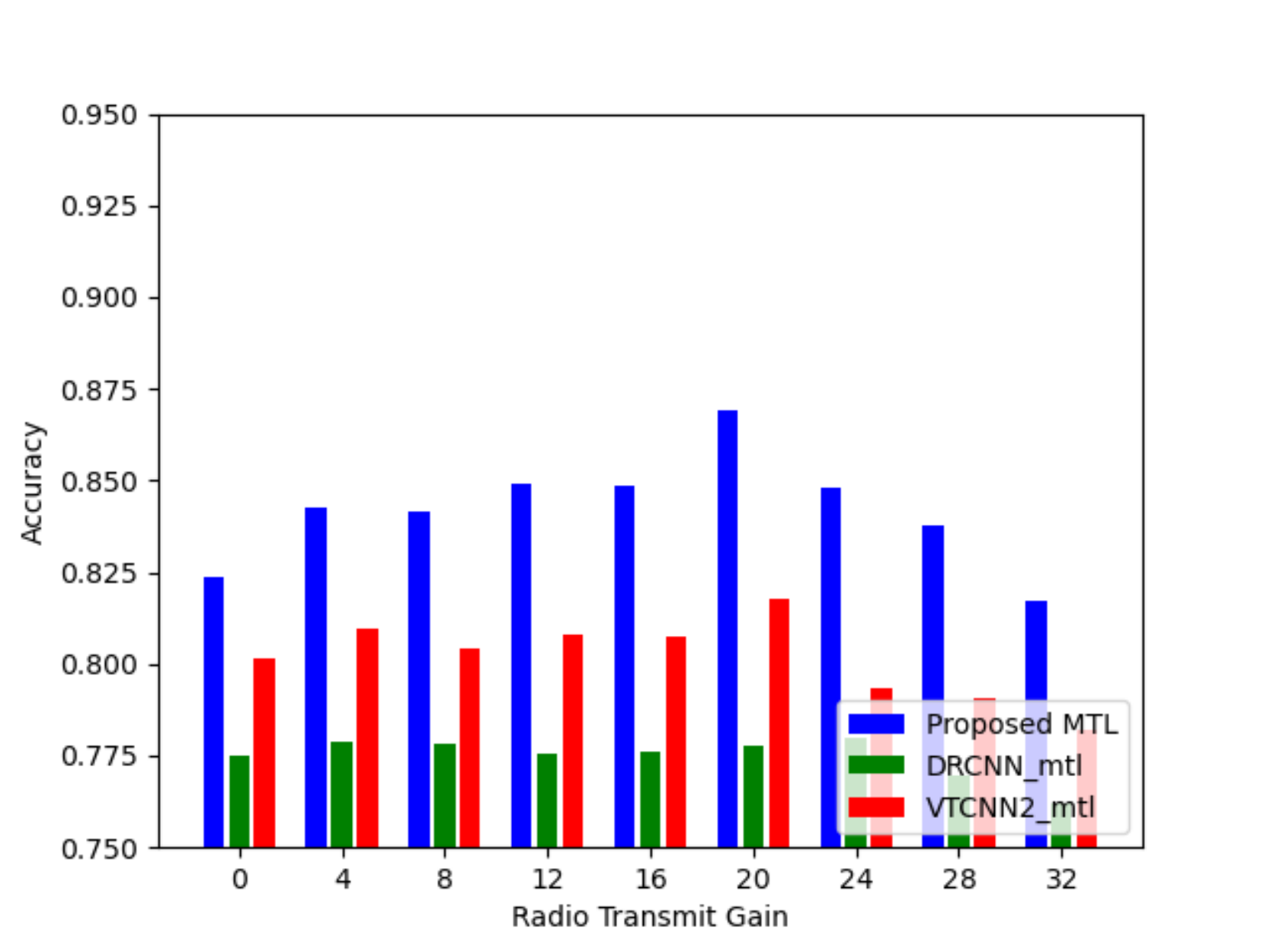} 
\caption{Modulation classification accuracy under varying OTA signal gains}
\label{fig:ota_mod}
\end{figure}

\begin{figure}
\centering
\includegraphics[width=0.99\columnwidth]{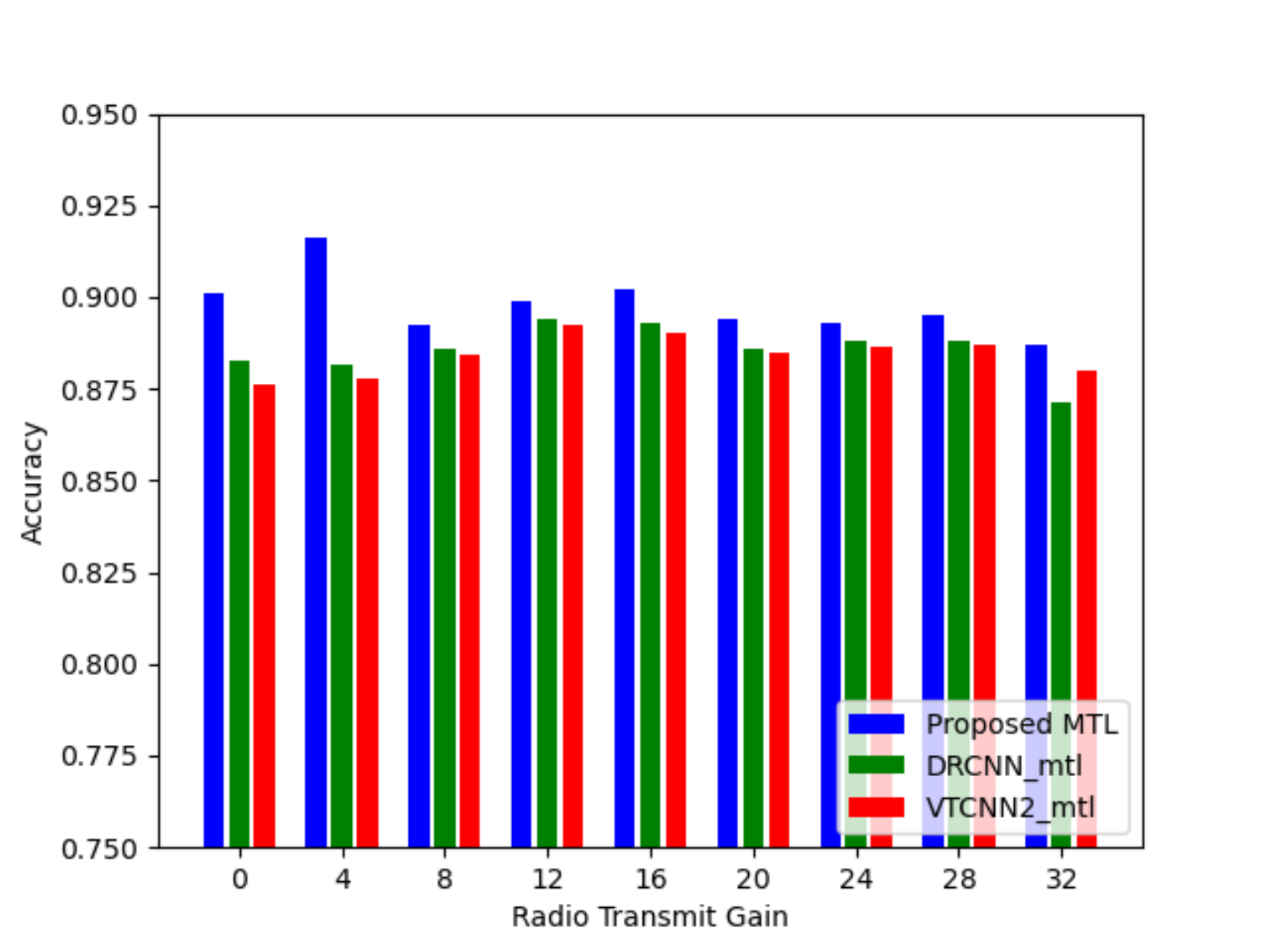} 
\caption{Signal classification accuracy under varying OTA signal gains}
\label{fig:ota_sig}
\end{figure}

With the above settings, we benchmarked the proposed MTL model with the VTCNN-MTL and DRCNN-MTL. Fig. \ref{fig:ota_mod} and Fig. \ref{fig:ota_sig} demonstrate the classification accuracy of the models under varying signal strengths. Please note here that the indoor laboratory restrictions constrained the radio separation causing the lowest transmitter gain itself to be of detectable signal strength. Increasing the transmitter gain further increased the noise floor which saturated the frontend of the receiver causing the classifier performance to slightly drop at higher transmitter gains. This trend is seen in all the three models that were evaluated with the proposed model being better than the other two on the same test data. Note here that the proposed MTL which was carefully designed to be lightweight in architecture since its inception outperforms the other two models in both modulation and signal classification tasks as shown in Table \ref{tab:bacc}. We note here that the higher overall signal classification accuracy for all three models can be attributed to the differences in the radar signals as shown in Fig. \ref{fig:9waves}.
These evaluations at varying signal powers validates the model performance in detecting and classifying even feeble signal (see Fig. \ref{fig:amp}) with its compact architecture. We emphasize here that the goal of this testbed evaluation was to demonstrate the applicability of MTL in learning multiple related signal characterization tasks jointly. 

\subsection{Computational and memory evaluation} 
\begin{table*}[h!]
    \centering
    \caption{Computational and Memory Evaluation \label{tab:comp}}
    \begin{tcolorbox}[tab2,tabularx={|p{3 cm}||p{3 cm}||p{3 cm}||p{3 cm}||p{3 cm}}]
      \textbf{Model}   & \textbf{FLOPs}  (Million) &\textbf{Parameters} (Million) &\textbf{CPU Inference time} (ms) & \textbf{Memory} (MB) \\ \hline\hline
      
      Proposed MTL model & \textbf{0.506} &\textbf{0.253} &\textbf{8.4} &\textbf{3.2}\\ \hline
      VTCNN-MTL &5.662 &2.83 &113.3 &34\\ \hline
      DRCNN-MTL &2.03 &1.01 &70.5 &12.2\\ \hline
    \end{tcolorbox}{}
\end{table*}

Having validated the performance of the proposed MTL model in terms of its top-1 classification accuracy in two simultaneous tasks - modulation and signal classification, we now evaluate the proposed model in terms of its computational and storage metrics. We resort to the following performance metrics to evaluate the computational and memory savings,
\begin{enumerate}

    \item \textbf{FLOPs} - The number of floating point operations in the model.
    \item \textbf{Parameters} - The number of trainable parameters in the model.
    \item \textbf{Memory} - The storage space required by the model in mega bytes (MB).
    \item \textbf{Inference time} - The time in seconds consumed by the model to generate an output for one instance of the 128 complex IQ samples input.
    
\end{enumerate}

Table \ref{tab:comp} clearly demonstrates the significant computational and memory savings of the proposed MTL model. The proposed architecture requires only 91.06\% and $\sim$75\% fewer FLOPs and trainable parameters in contrast to the MTL versions of VTCNN and DRCNN respectively. The lightweight MTL performs faster inferences at the rate of 8.4ms on an Intel Core i5-3230M CPU, consuming 90.5\% and 73.8\% lesser memory requirement in contrast to VTCNN-MTL and DRCNN-MTL respectively. These evaluations further validate the applicability of the proposed MTL model for the RF edge applications.

\section{Compressed Model - Quantized Neural Network}
As one of the key motivations of this paper is to enable and demonstrate the design of lightweight neural network architecture for support on resource-constrained edge devices, we now discuss the prospects of model compression. The challenging part of the MTL design process is the careful hyper-parameter tuning as discussed in section \ref{sec:analysis} to maintain the classification accuracy across multiple tasks. In this section, we emphasize the significance of model compression and present it as an important step to consider in the deep learning pipeline for resource-limited edge platforms.

\begin{figure}[h!]
    \centering
\includegraphics[width=0.9\columnwidth]{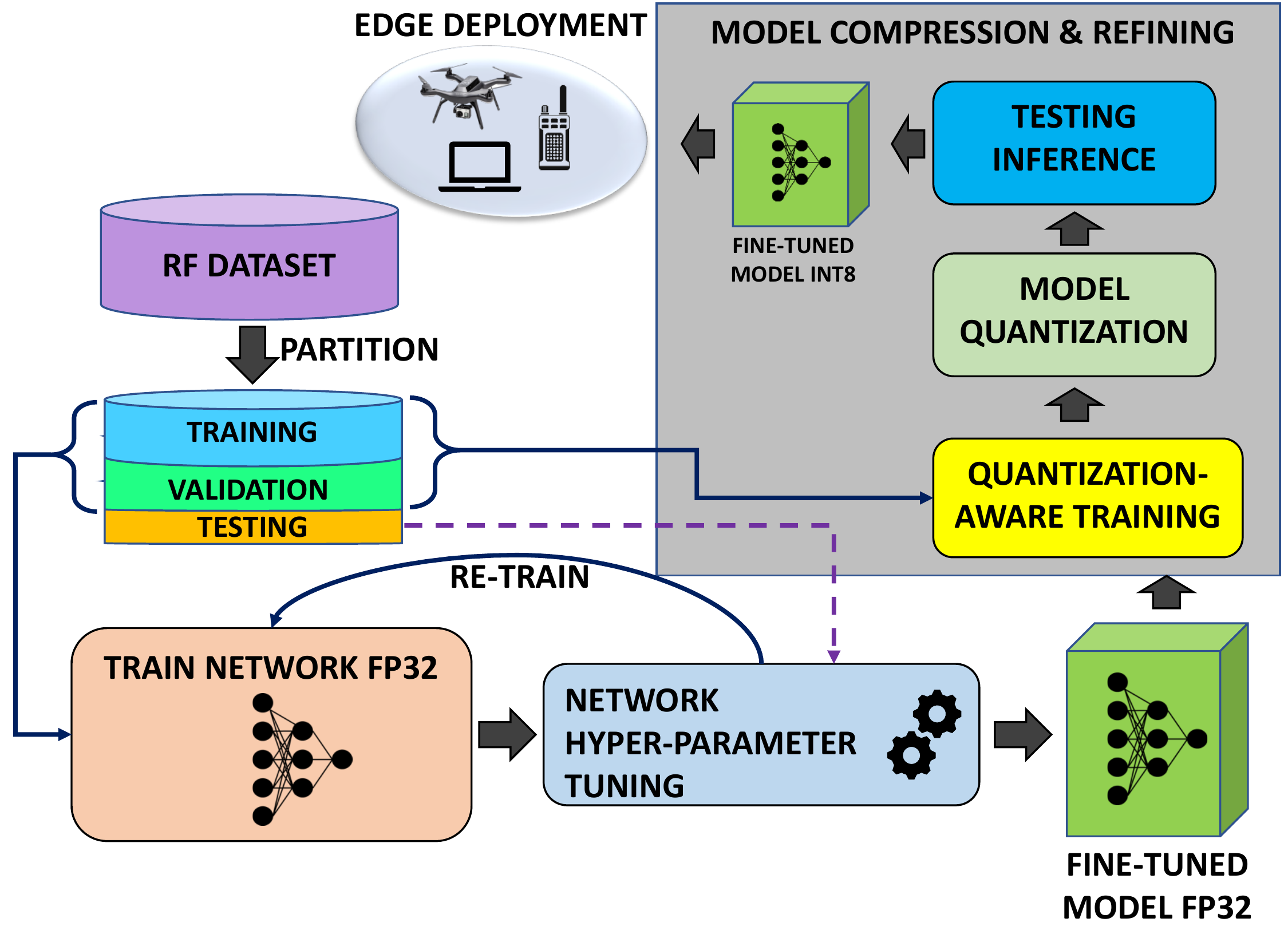} 
\caption{Deep Learning Model Training and Deployment pipeline.}
\label{fig:pipe}    
\end{figure}

Fig. \ref{fig:pipe} shows the proposed deep learning pipeline whereby the model is trained, validated, tested, and also undergoes a necessary step \emph{model compression and refining} to arrive at a well balanced compressed model. Although model compression techniques have been explored and practiced heavily for computer  vision and NLP, their application in the wireless realm is rudimentary. Hence, we present this design methodology to provide interested readers valuable firsthand insight and demonstrate the performance of the compressed MTL model. We further emphasize that the lack of information regarding the \emph{howto} of model compression or its effect on the crafted model from a wireless communication standpoint acts as a barrier and thwarts active research in this direction. 

\begin{figure}
\centering
\includegraphics[width=0.99\columnwidth]{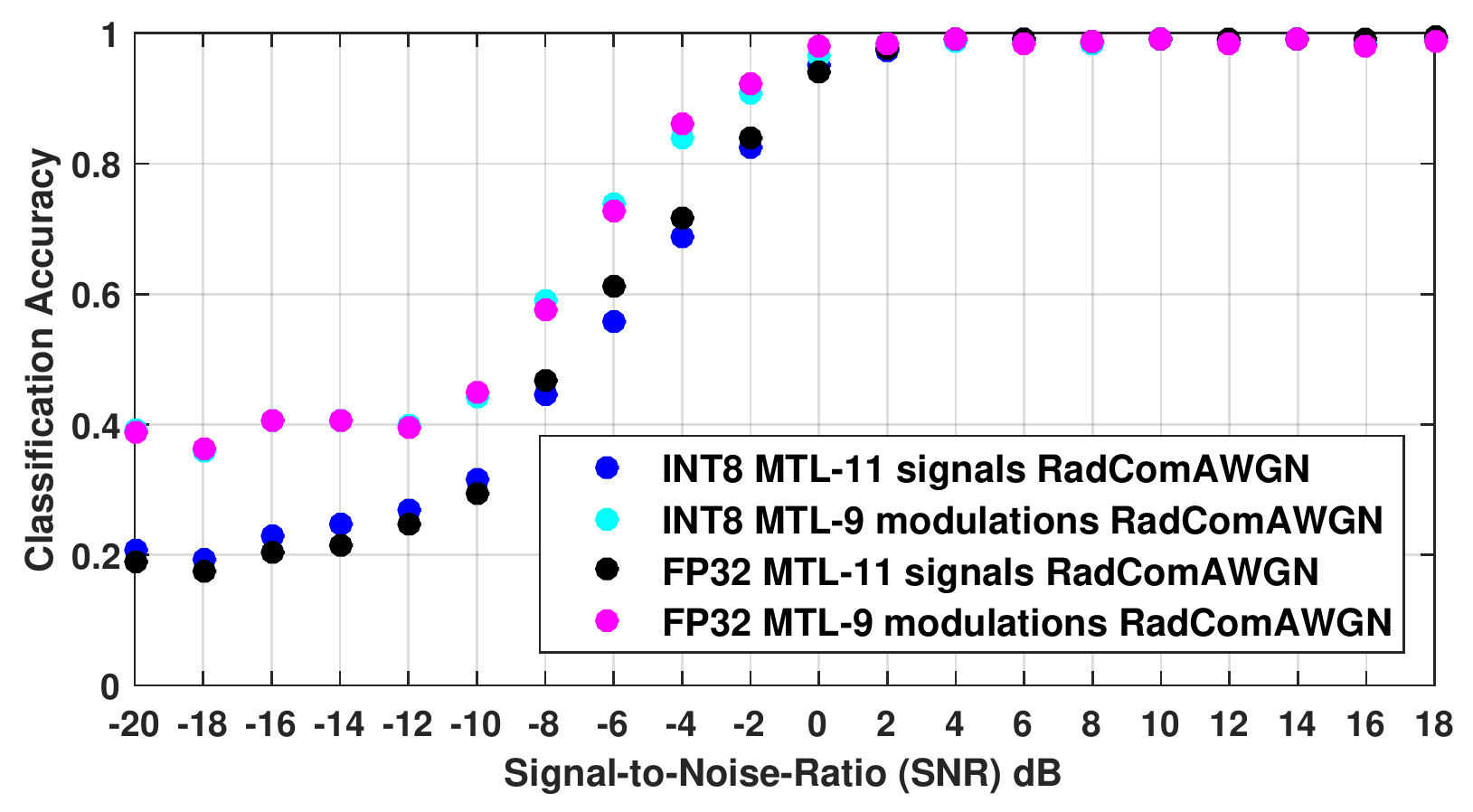} 
\caption{MTL classification under varying noise}
\label{fig:int8_snr}    
\end{figure}

Deep learning models, in general, consume substantial compute and memory resources \cite{datacenter_fb}, which can exhaust even powerful servers let alone resource-constrained edge devices. Several algorithms, software, and hardware have been proposed and implemented to lighten the imposed burden on resources \cite{datacenter_fb,quantize}. Among which, \emph{quantization} is a promising approach. Quantization is a reduced precision strategy whereby the weights and activations of the trained neural network is shrunk from higher to lower bit precision. The standard number format in deep learning is floating point 32 (FP32), specified in IEEE754 \cite{ieee754}. The FP32 representation is referred to as single precision. Similarly, the format which use half the bits of FP32 is called half precision (floating point 16). In this paper, we will quantize the MTL to 8 bit integer (INT8) \emph{reducing the memory footprint by a factor of four} \cite{int8_1,int8_2}. This is a most used form of quantization whereby a FP32 tensor $\mathbf{F}_{32}$ is represented by a INT8 tensor $\mathbf{I}_8$ along with auxiliary attributes - scale $s$ and zero point $z$ as in equation \ref{eq:quant}.
\begin{equation}
    \mathbf{F}_{32} = s*(\mathbf{I}_8 - z )
    \label{eq:quant}
\end{equation}

In our empirical evaluations, we have found quantized model consumes fewer memory and computing resources while keeping its accuracy close to the unquantized model. Prior works adopting INT8 quantization too have demonstrated this model accuracy preservation \cite{int8_1,int8_2}. Additionally, we chose INT8 quantization partly due to the fact that the widely accepted computational platforms such as ARM CPUs, Intel CPUs, and NVIDIA GPUs are introducing low-level instructions set to support INT8 computation efficiently. Several deep learning frameworks are available today - TensorFlow Lite, PyTorch, MXNet to perform model quantization. 

In this paper, we used the TensorFlow Lite framework \cite{tflite} since our model was implemented in Keras with TensorFlow backend. TensorFlow Lite is part of the TensorFlow library and is intended to support model deployment on mobile and IoT edge devices.
Quantization schemes can be broadly divided into two categories - post-training quantization and quantization aware training. Post-training quantization starts with a trained FP32 model and performs calibration on a cross-validation dataset to find the best quantization parameters. On the other hand, quantization aware training models quantization during training and can provide higher accuracies in contrast to post-training quantization. For this reason, we chose the quantization aware training method. Below, we summarize the steps required to attain an INT8 quantized model leveraging the TensorFlow Lite framework,
\begin{enumerate}
    \item Train a FP32 model and fine tune without quantization aware training. This is the FP32 finetuning discussed in the previous sections.
    \item Apply quantization aware training to the whole model. For this, use the \texttt{keras.quantize\_model} function from the \texttt{tensorflow\_model\_optimization} library. Compile the model and train as usual in Keras. Let the trained model be denoted as QatModel
    \item Next, we create a quantized model with INT8 weights by loading the QatModel to the converter class as \texttt{tf.lite.TFLiteConverter.from\_keras\_\\model(QatModel)}. 
    \item Specify the optimization policy for the converter class as \texttt{tf.lite.Optimize.DEFAULT} which would quantize the weights to 8-bits precision. Finally, the \texttt{convert} function will generate the quantized model.
\end{enumerate}
 We state here that the FP32 model which in itself was handcrafted to be lighter in architecture had a model size of 2.97 MB. The INT8 quantization yielded a $11.8\times$ smaller model, i.e., of size 251.6 kB. 

 Fig. \ref{fig:int8_snr} shows the classification accuracy comparison of the INT8 model with its unquantized FP32 counterpart on the RadComAWGN dataset. It can be seen that the INT8 quantization resulted in almost no accuracy loss giving a 98.9\% and 99.2\% signal and modulation classification accuracy respectively at 10 dB and over 98\% accuracy (modulation and signal classification) at 0 dB and above. Recall here that the INT8 model is $11.8\times$ smaller than the FP32 model and performs as good while reducing the memory and computational load.
 
 With these evaluations, we would like to summarize that the quantized MTL model which is only 251.6 kB can perform two waveform characterization tasks while attaining a very high accuracy for signals even at 0 dB.
 



\section{Conclusion and Future Work}
The key novelty of our paper lies in proposing a multi-task learning framework for solving two challenging waveform characterization tasks. The proposed framework lays special emphasis on designing lighter architecture tailored for resource-limited IoT platforms. We present a walkthrough of the lighter architecture design methodology and demonstrated the applicability of the MTL framework in performing related signal characterization tasks jointly on synthetic as well as OTA dataset. Specifically, we demonstrate that the proposed architecture requires only 91.06\% and $\sim$75\% fewer FLOPs and trainable parameters in contrast to the MTL versions of VTCNN and DRCNN respectively. The proposed lightweight MTL performs faster inferences at the rate of 8.4ms on an Intel Core i5-3230M CPU, consuming 90.5\% and 73.8\% lesser memory requirement in contrast to VTCNN-MTL and DRCNN-MTL respectively. 

Further, we present a deep learning pipeline tailored for beyond 5G IoT frameworks alongside a slice of neural network quantization. The $11.8\times$ compressed model was able to perform as good as the unquantized counterpart with very negligible accuracy loss. We advocate the adoption of such MTL frameworks for the future communication networks to ease resource-burden by enabling a single model to perform multiple tasks. The feasibility established by the proposed MTL architecture provided incentive for the future efforts in this domain to extend the model to include more tasks such as emitter classification, sampling rate regression, bandwidth regression, among others. The inclusion of additional signal characterization tasks will be part of our future research. Finally, the RadComOta dataset comprising radar and communication waveforms (collected under OTA setting) that can be used for modulation and/or signal classification tasks has been made publicly available to promote future research \cite{data}. 

As part of our future research efforts, we hope to extend the dataset further to include other signal types and characteristics. Additionally, we intend to closely evaluate the effect of various structured pruning approaches with an in-depth review of saliency of each filter in the convolutional layer for efficient hardware acceleration.
\bibliographystyle{ieeetr}
\bibliography{bibfile1}


\begin{IEEEbiography}[{\includegraphics[width=1in,height=1.25in,clip,keepaspectratio]{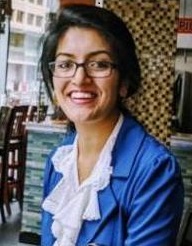}}]{Anu Jagannath} currently serves as the Founding Associate Director of Marconi-Rosenblatt AI/ML Innovation Lab at ANDRO Computational Solutions, LLC. She received her MS degree from State University of New York at Buffalo in Electrical Engineering. She is also a part-time PhD candidate with the Dept. of Electrical and Computer Engineering at Northeastern University, USA. Her research focuses on MIMO communications, Deep Machine Learning, Reinforcement Learning, Adaptive signal processing, Software Defined Radios, spectrum sensing, adaptive Physical layer, and cross layer techniques, medium access control and routing protocols, underwater wireless sensor networks, and signal intelligence. She has rendered her reviewing service for several leading IEEE conferences and Journals. She is the co-Principal Investigator (co-PI) and Technical Lead in multiple Rapid Innovation Fund (RIF) and SBIR/STTR efforts involving  applied AI/ML for wireless communications. She is also the inventor on 6 US Patents (granted and pending).
\end{IEEEbiography}

\begin{IEEEbiography}[{\includegraphics[width=1in,height=1.5in,keepaspectratio]{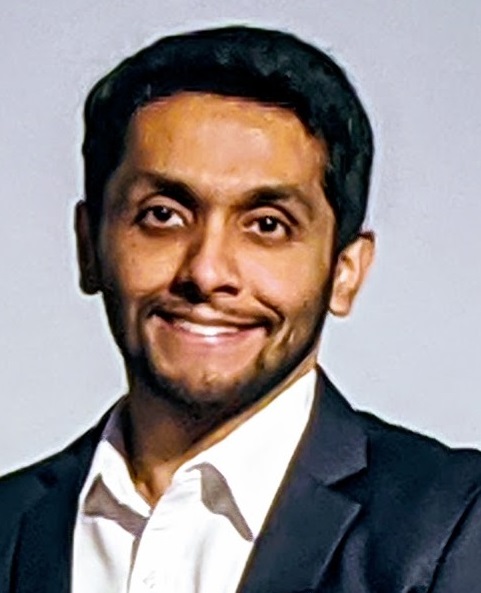}}]{Jithin Jagannath} (SM'19) is the Chief Scientist of Technology and Founding Director of the Marconi-Rosenblatt AI/ML Innovation Lab at ANDRO Computational Solutions. He is also the Adjunct Assistant Professor in the Department of Electrical Engineering at the University at Buffalo, State University of New York. Dr. Jagannath received his B. Tech in Electronics and Communication from Kerala University;  M.S. degree in Electrical Engineering from University at Buffalo, The State University of New York; and received his Ph.D. degree in Electrical Engineering from Northeastern University. Dr Jagannath was the recipient of 2021 IEEE Region 1 Technological Innovation Award with the citation, "For innovative contributions in machine learning techniques for the wireless domain''. He is also the recipient of AFCEA International Meritorious Rising Star Award for achievement in Engineering and AFCEA 40 Under 40 in 2022. 

Dr. Jagannath heads several of the ANDRO's research and development projects in the field of Beyond 5G, signal processing, RF signal intelligence, cognitive radio, cross-layer ad-hoc networks, Internet-of-Things, AI-enabled wireless, and machine learning. He has been the lead and Principal Investigator (PI) of several multi-million dollar research projects. This includes a Rapid Innovation Fund (RIF) and several Small Business Innovation Research (SBIR)s for several customers including the U.S. Army, U.S Navy, Department of Homeland Security (DHS),  United States Special Operations Command (SOCOM). He is currently leading several teams developing commercial products such as SPEARLink\texttrademark, DEEPSpec\texttrademark~ among others. He is an IEEE Senior member and serves as IEEE Industry DSP Technology Standing Committee member. Dr. Jagannath's recent research has led to several peer-reviewed journal and conference publications. He is the inventor of Ten U.S. Patents (granted, pending, and provisional). He has been invited to give various talks including Keynote on the topic of machine learning and Beyond 5G wireless communication. He has been invited to serve on the Technical Program Committee for several leading technical conferences.
\end{IEEEbiography}

\end{document}